\documentclass[12pt]{article}
\usepackage{epsfig}

\def\la{\mathrel{\mathpalette\fun <}}

\def\fun#1#2{\lower3.6pt\vbox{\baselineskip0pt\lineskip.9pt
\ialign{$\mathsurround=0pt#1\hfil##\hfil$\crcr#2\crcr\sim\crcr}}}
\newcommand{\bea}{\begin{eqnarray}}
\newcommand{\eea}{\end{eqnarray}}
\newcommand{\nn}{\nonumber}
\newcommand{\beq}{\begin{equation}}
\newcommand{\eeq}{\end{equation}}

\begin{document}

\title{Colour effective particles and confinement
}
\author{ V.V. Anisovich\\
{\footnotesize PNPI, Gatchina 188300, Russia} }
\date{\today}
\maketitle

\begin{abstract}

 \centerline{\it Talk given  at the Workshop ''Hadron Structure and QCD'',
 Gatchina, Russia, 3-6 July 2010}
Here I present a
 brief review of papers where
 the idea is pushed forward  that colour confinement is realized by singular
interaction
at large distances between
colour effective particles (constituent quarks, diquarks, massive effective
gluons).

\end{abstract}

 \section {Constituent quarks as colour effective particles}

The first successful steps in  understanding the  internal
structure of hadrons were made
in 60's by introducing an idea of constituent quarks, according to which
baryons are
three-quark systems and mesons are quark-antiquark ones.
Non-relativistic description of
these systems gave rise to the systematization of low-lying states
($SU(6)$ symmetry) and allowed one to work with  constituent quarks as
 non-flying out objects due to the potential barrier. The weak
point of this approach is its inapplicability to highly excited states.
Later on, experimental and theoretical studies revealed  more
complicated structure of hadrons,  constituent quarks being
spatially separated clusters of the QCD particles. It allows us to
consider the constituent quark as an effective particle, {\it i.e.}
 a ``dressed quark'' of the valence QCD-quark.

\begin{figure}[hbt]
%Fig 1
\centerline{\epsfig{file=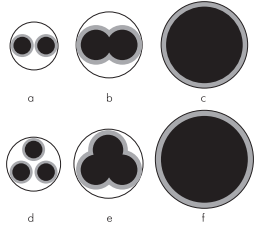,width=1.3in}}
\caption{ Quark structure of a meson (a--c) and a baryon (d--e) in
the constituent quark model. At moderately high energies (a,d) constituent quarks
inside the hadron are spatially separated. With the energy growth,
quarks overlap partly  (b,e); at superhigh energies
(c,f) quarks are completely overlapped, and hadron--hadron collisions
loose the property of additivity.  }\label{Y1f11}
\end{figure}

For reactions at low
and moderately high energies the masses of the light constituent quarks
( $q=u,\,d,\,s$) are of the order of
 $m_u\simeq m_d\simeq 300-400$ MeV, $m_s\simeq  500$ MeV.
The size of the light constituent quark
 $ \langle r^2_{constituent \,
quark}\rangle \simeq 0.1$ fm$^2$.

\section{Systematization of mesons in terms of
constituent quarks}

\begin{figure}[hbt]
%Fig.2
\centerline{\epsfig{file=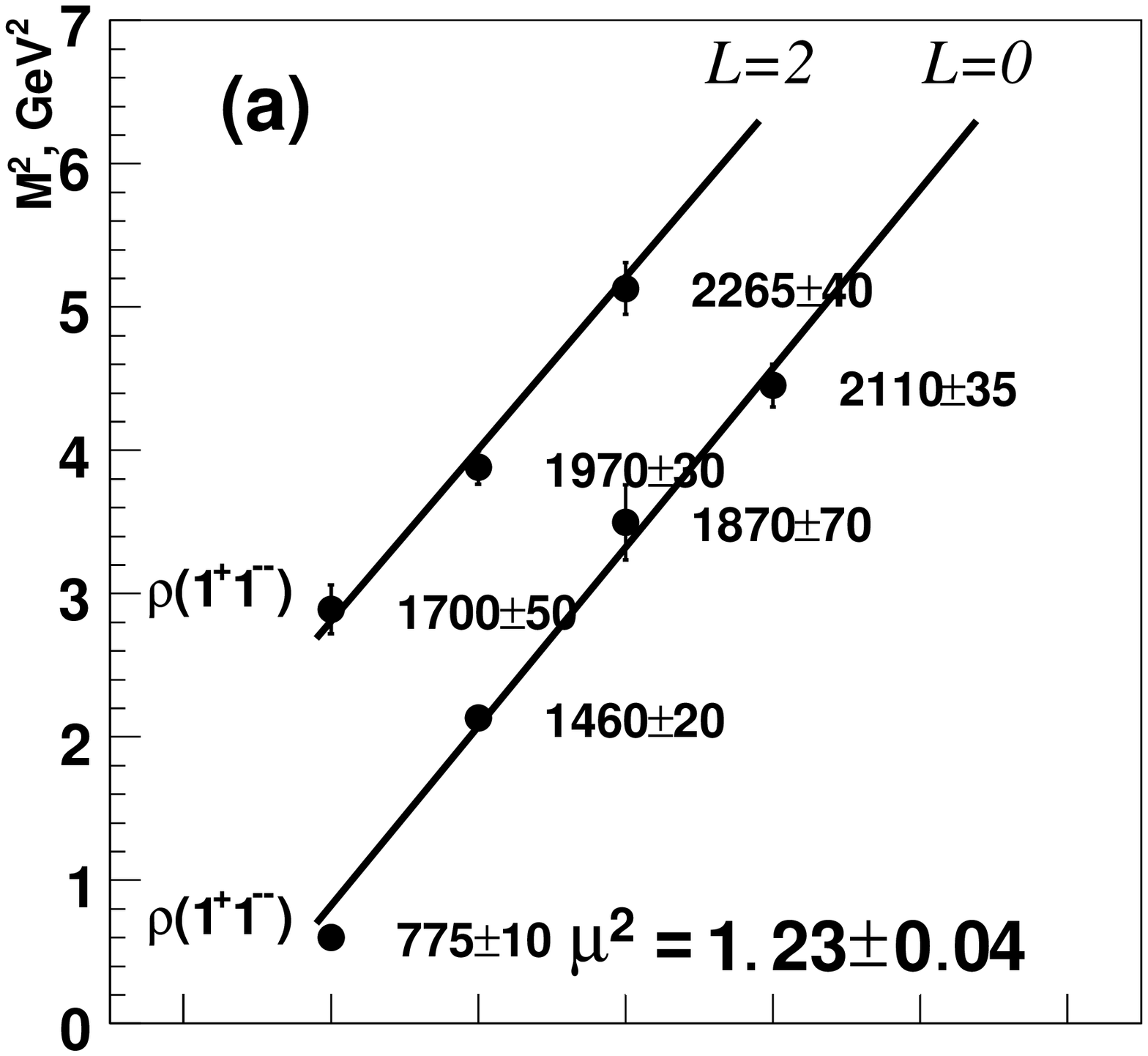,width=6.0cm}
\hspace{-5mm}
            \epsfig{file=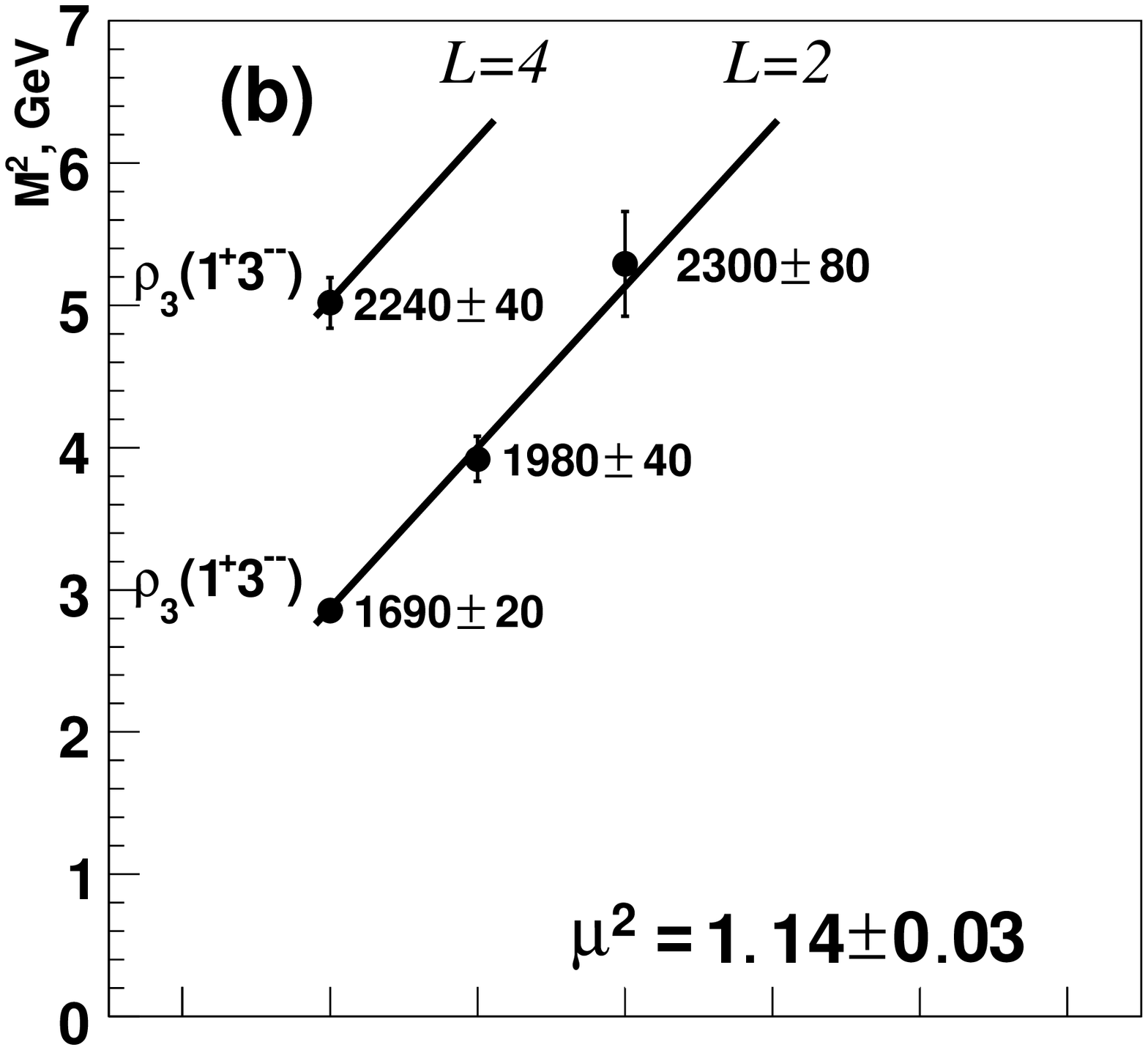,width=6.0cm}}
%\vspace{-4.95mm}
\caption{Trajectories for $(C=-)$ meson states on the $(n,M^2)$
plane. The black dots mark the observed states, open circles stand for
the predicted states.
 For mesons, the used notation is
$\quad (I^G J^P)$. }\label{2}
\end{figure}

\begin{figure}[hbt]
 %Fig.3
\centerline{\epsfig{file=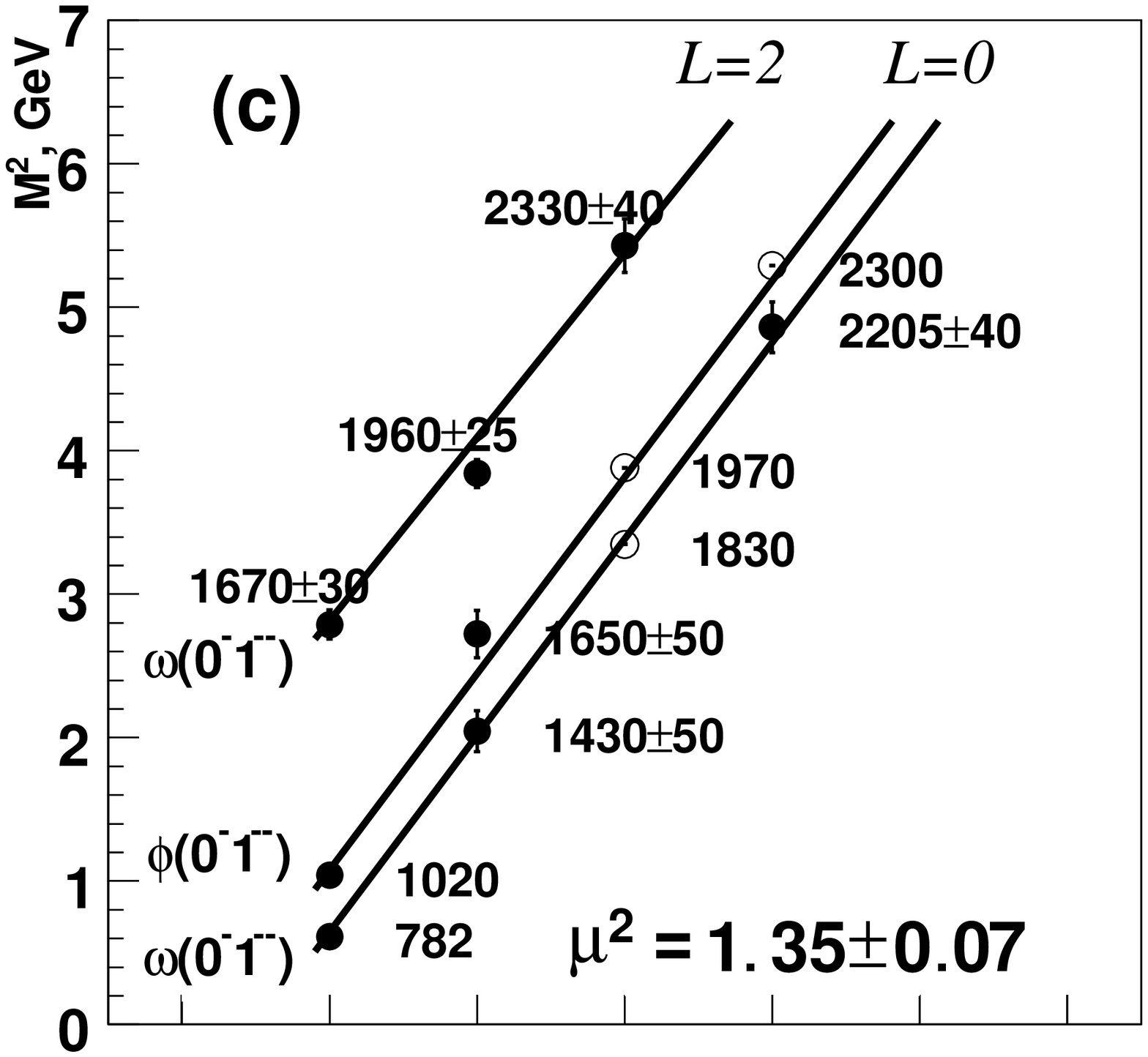,width=6.0cm}\hspace{-5mm}
            \epsfig{file=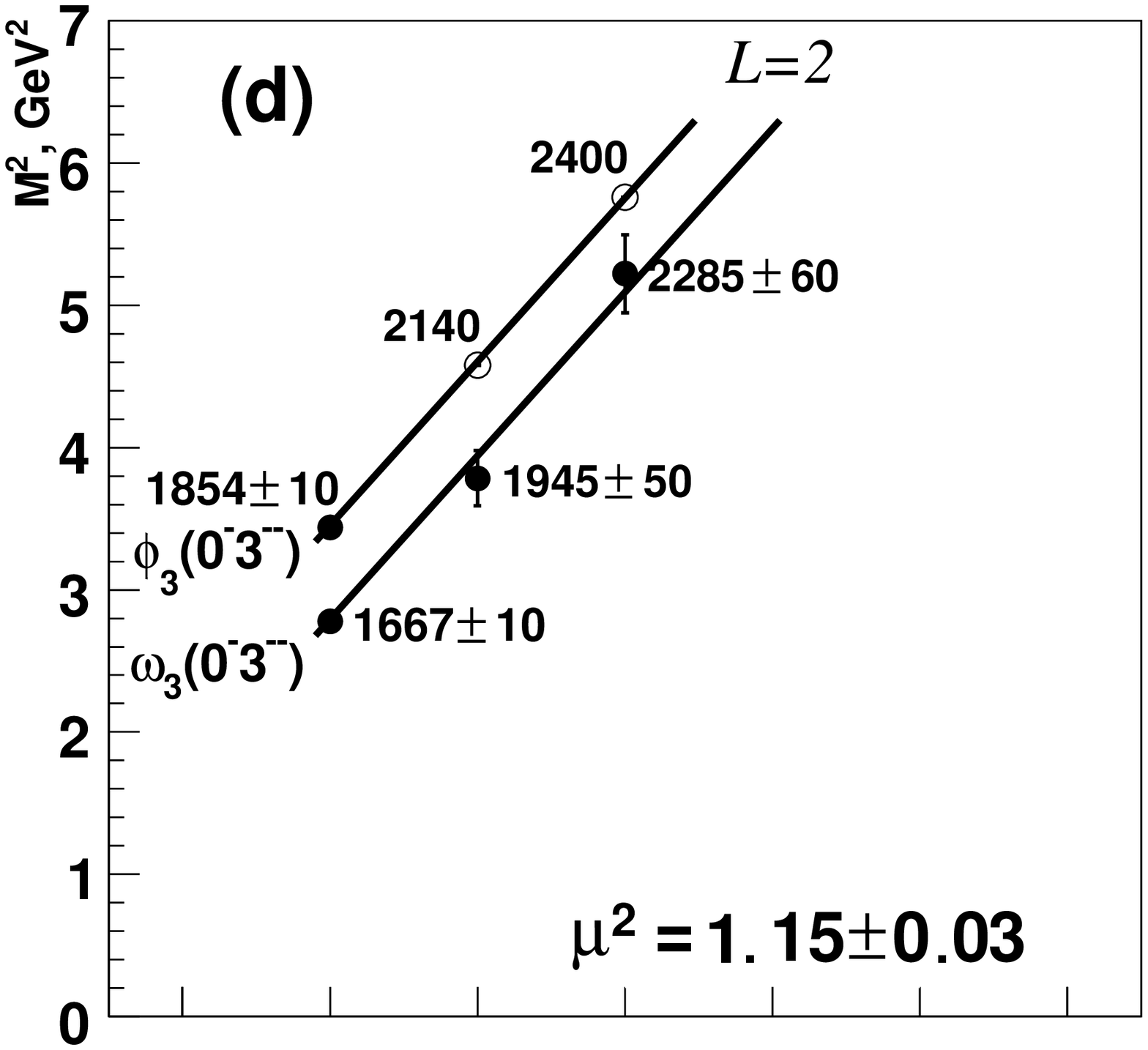,width=6.0cm}}
%\vspace{-4.95mm}
\caption{Trajectories for $(C=-)$ meson states. Isoscalar states have two flavour
components each, $n\bar n=(u\bar{u}+ d\bar{d})/\sqrt{2}$ and $s\bar{s}$, this doubles the
number of trajectories. }
 \label{3}
 \end{figure}

\begin{figure}[hbt]
%Fig.4
\centerline{\epsfig{file=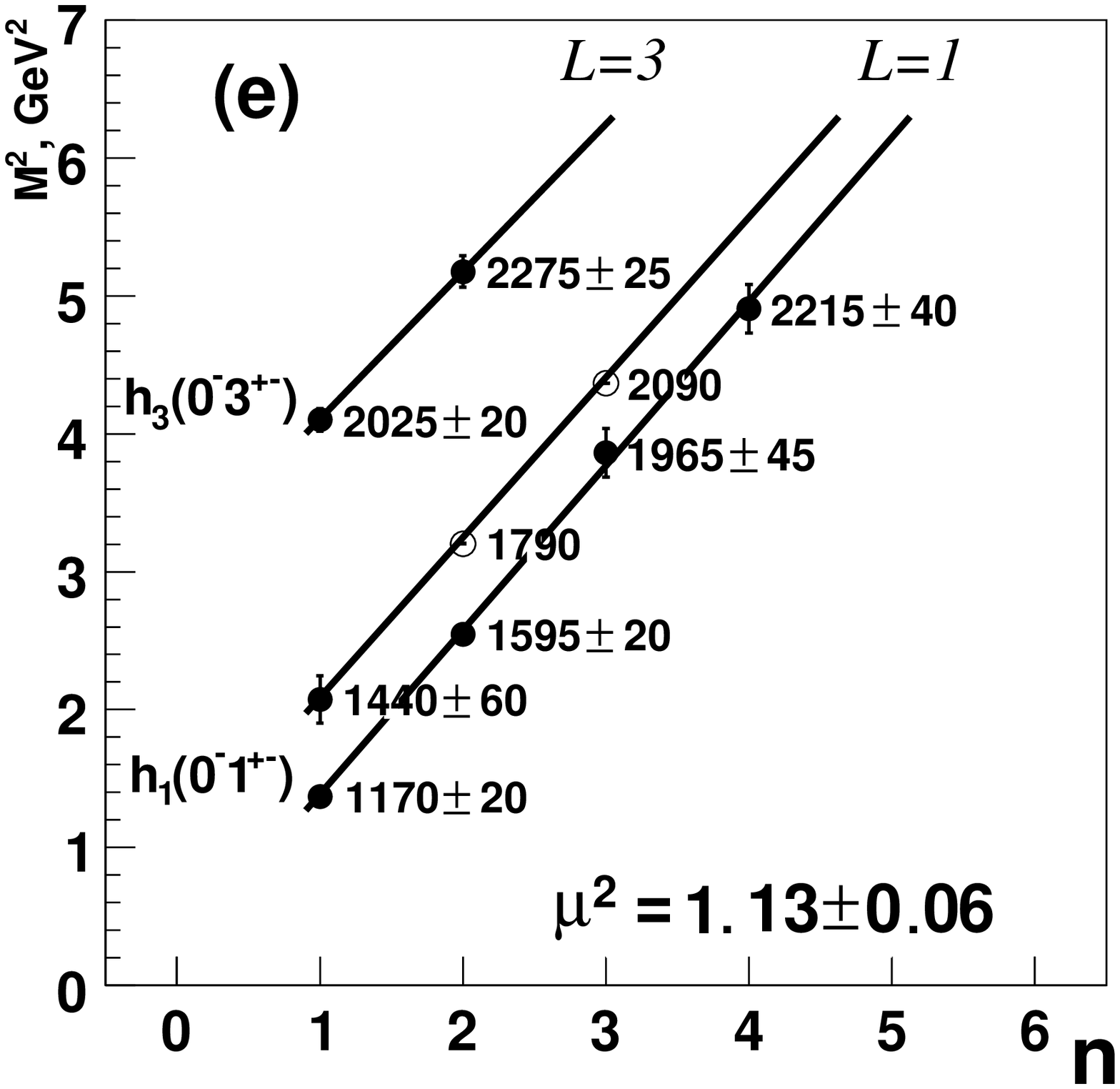,width=6.0cm}\hspace{-5mm}
            \epsfig{file=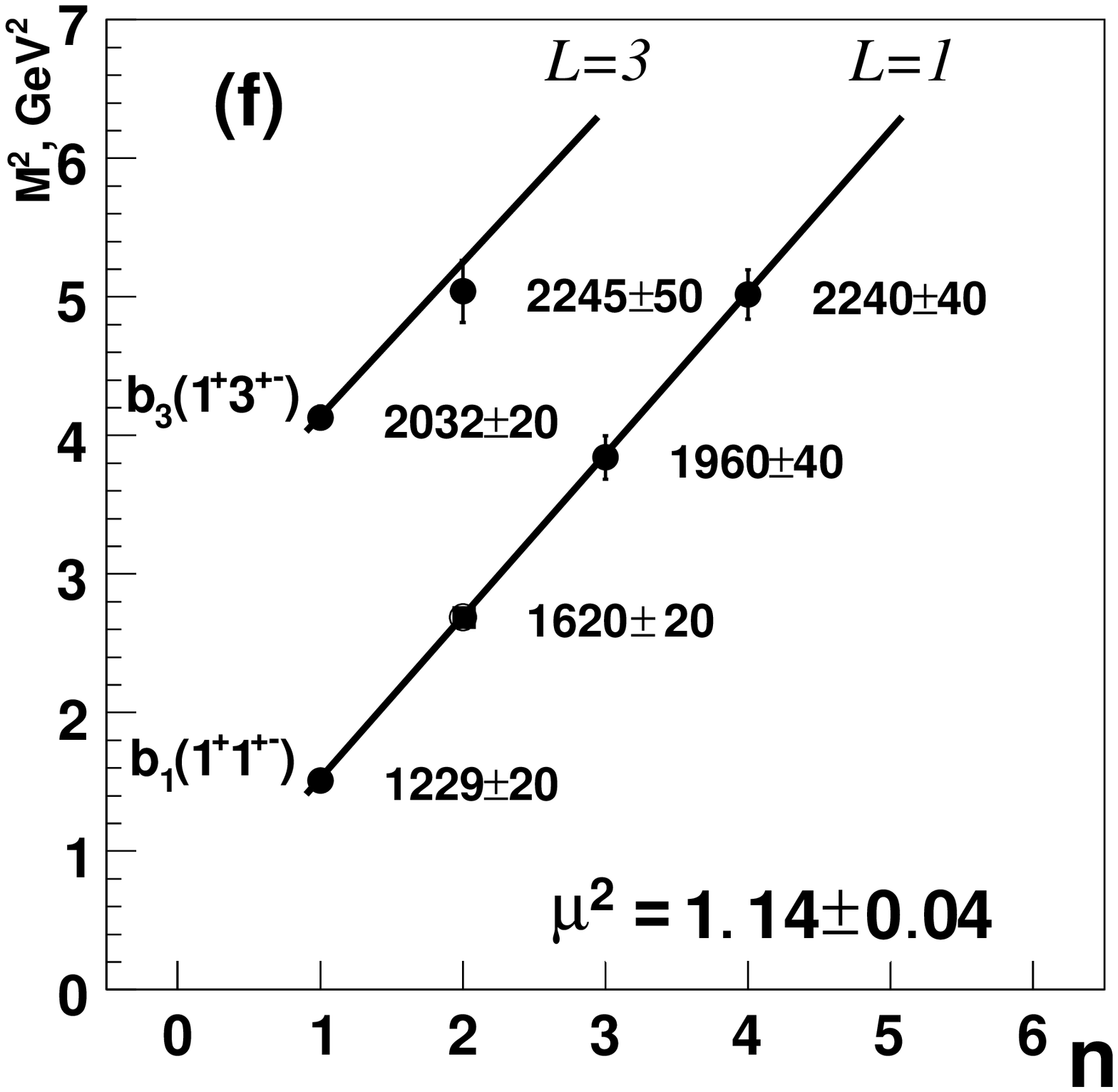,width=6.0cm}}
\vspace{-2.00mm}
\caption{Trajectories for $(C=-)$ meson states. States
with $J=L\pm 1$ have two components: at fixed $J$ there are states with
$L-1$ and $L+1$, so one may assert the doubling of trajectories at fixed
$J$, for example,    for
 $(1,1^{--})$ and $(1,3^{--})$.}
\label{4}
 \end{figure}

The systematization of mesons support the idea of  constituent quarks.
 A decade ago,  considerable progress was reached in the
determination of highly excited meson states in the mass region
1950--2400~MeV. These results allowed us to
systematize  $q\bar{q}$-meson states on the  $(n,M^2)$ and
$(J,M^2)$ planes, $n$ being the radial quantum number of a
$q\bar{q}$ system with mass $M$ and spin $J$. The $q\bar{q}$ states,
$n^{2S+1}L_Jq\bar {q}$,
fill in the
following  $(n,M^2)$ trajectories:
\begin{eqnarray}
\label{X2e11}
& & ^1S_0\ \to\ \pi(10^{-+}), \ \eta(00^{-+})\ ; \nonumber\\
& & ^3S_1\ \to\ \rho(11^{--}),\ \omega(01^{--})/\phi(01^{--})\ ;
\nonumber\\
& & ^1P_1\ \to\ b_1(11^{+-}),\ h_1(01^{+-})\ ; \nonumber\\
& & ^3P_J\ \to\ a_J(1J^{++}),\ f_J(0J^{++}),\ J=0,1,2\ ;\nonumber\\
& & ^1D_2\ \to\ \pi_2(12^{-+}),\ \eta_2(02^{-+})\ ;\nonumber\\
& & ^3D_J\ \to\ \rho_J(1J^{--}),\ \omega_J(0J^{--})/\phi_J(0J^{--}),\
J=1,2,3\ ; \nonumber\\
& & ^1F_3\ \to\ b_3(13^{+-}),\ h_3(03^{+-})\ ;\nonumber\\
& & ^3F_J\ \to\ a_J(1J^{++}),\ f_J(0J^{++}),\  J=2,3,4\ .
\end{eqnarray}
Trajectories for $(C=-)$ meson states on the $(n,M^2)$
plane are shown in Figs. \ref{2}, \ref{3}, \ref{4}.

Recall that $L$ refers to the quark notation of state: $n^{2S+1}L_Jq\bar q$.
Linear trajectories give the $n^{2S+1}L_Jq\bar q$ states:
\beq
M^2=M^2_0+(n-1)\mu^2,
\eeq
where $M_0$ is the mass of the basic state, $n=1$ . The slope $\mu^2$ is of the same
order for the all trajectories.

The linearity of  $(n,M^2)$ trajectories at $M\la 2.4$ GeV with the universal slope
$\mu^2\simeq 1.15$ GeV$^2$ was observed in \cite{syst} -- in the region of large masses it
was essentially based on meson spectra analyses in the $p\bar p$
annihilation \cite{PNPI-RAL},
partial wave analysis of $\pi^+ \pi^- \pi^0$ production in two-photon collisions at
LEP \cite{Sh}
 and simultaneous $K$-matrix fit of a number of meson
spectra  \cite{km1}. Recently performed $K$-matrix analyses \cite{km2} confirm the
linearity of  trajectories.

\begin{figure}
%Fig. 5
\centerline{\epsfig{file=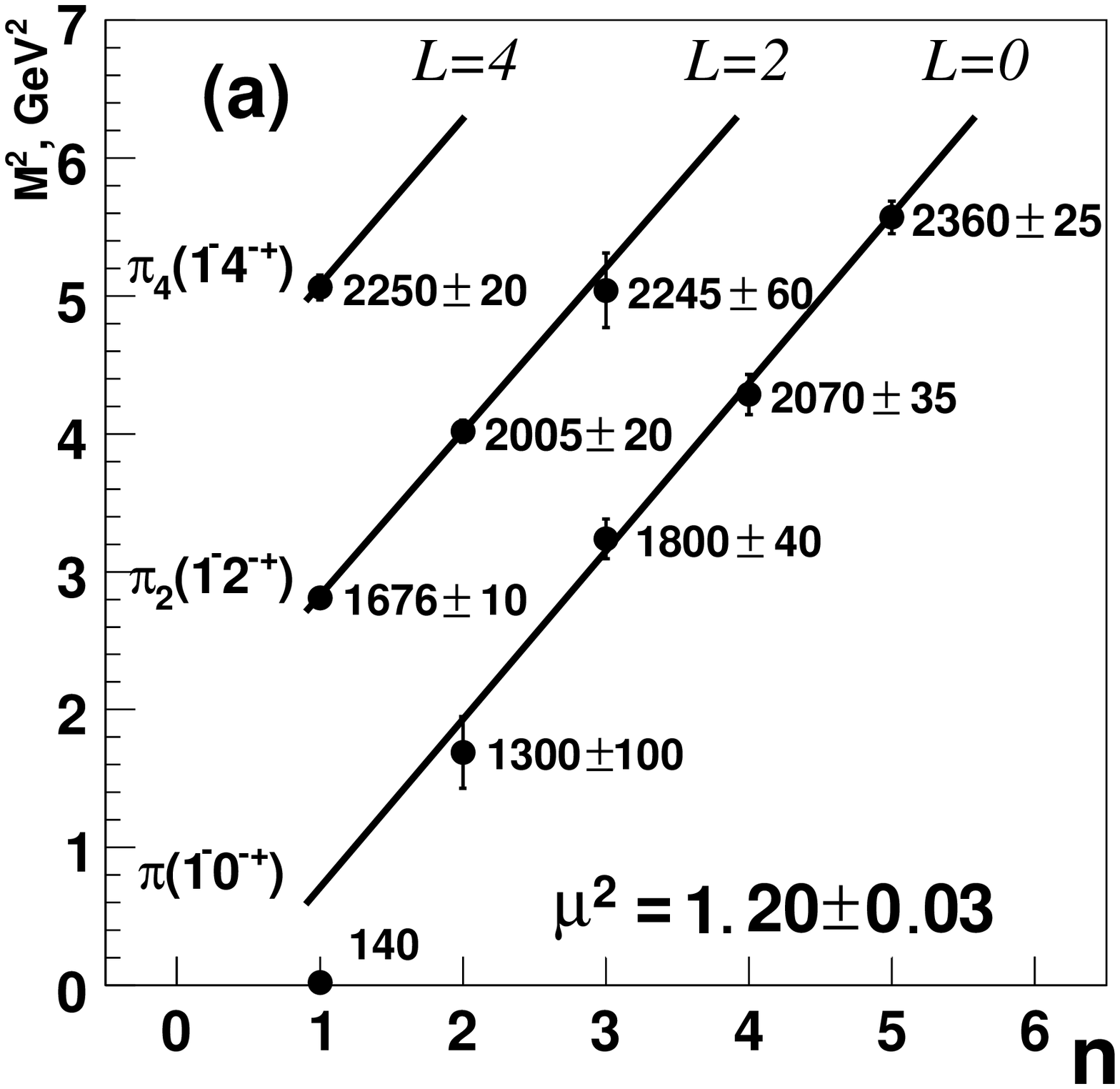,width=7.0cm}\hspace{-5mm}
            \epsfig{file=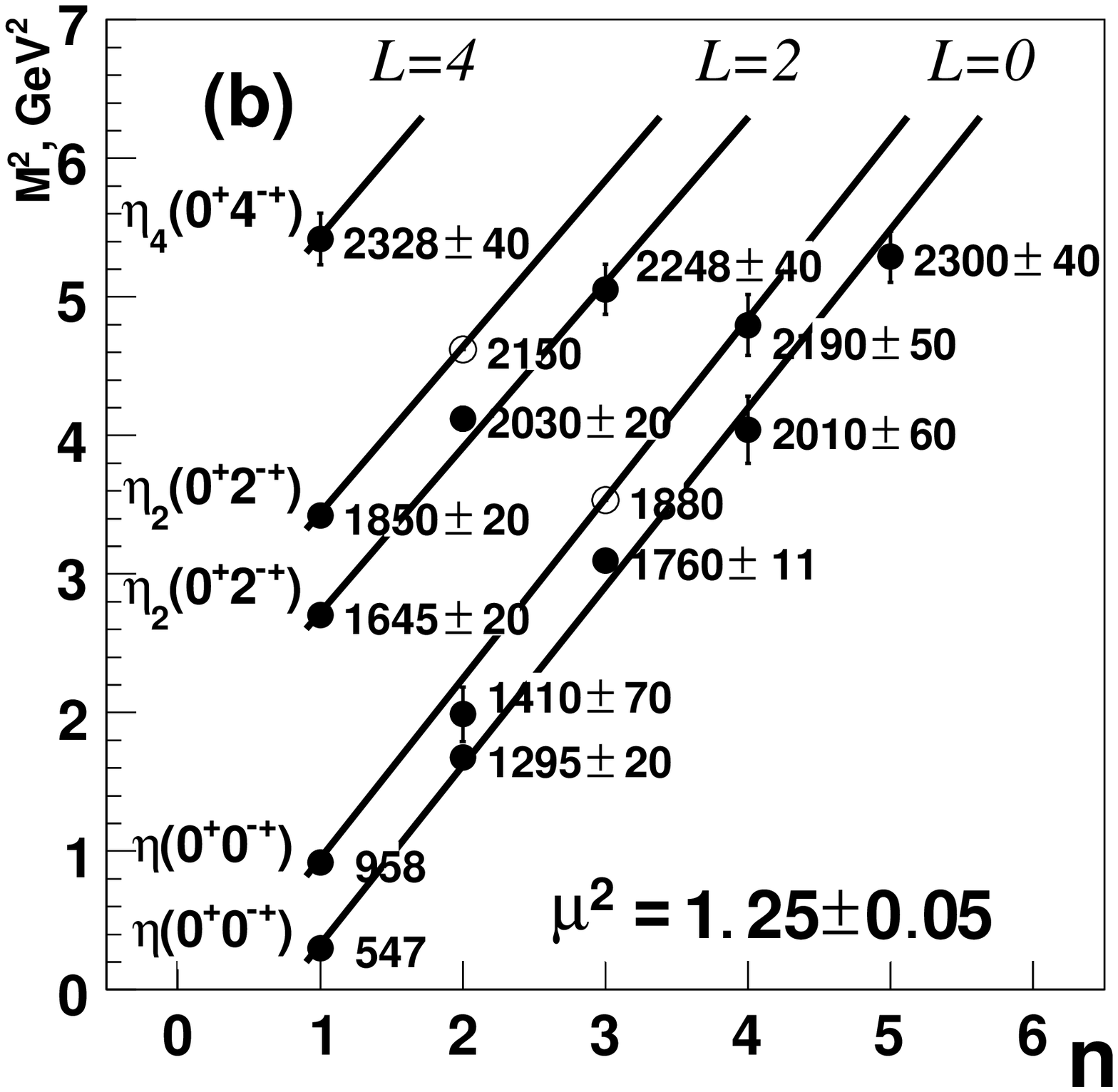,width=7.0cm}}
\vspace{-7.0mm}
\centerline{\epsfig{file=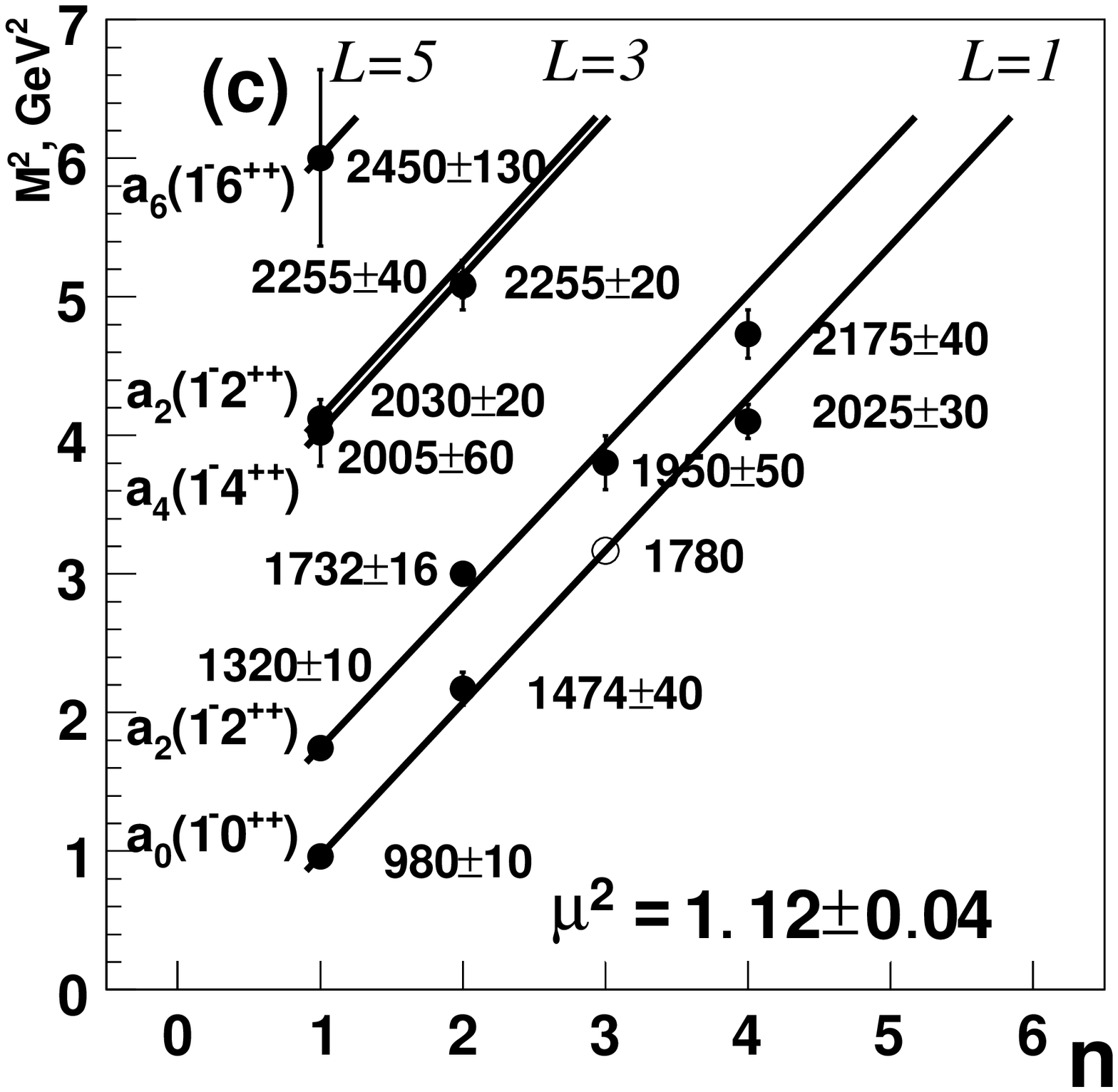,width=7.0cm}\hspace{-5mm}
            \epsfig{file=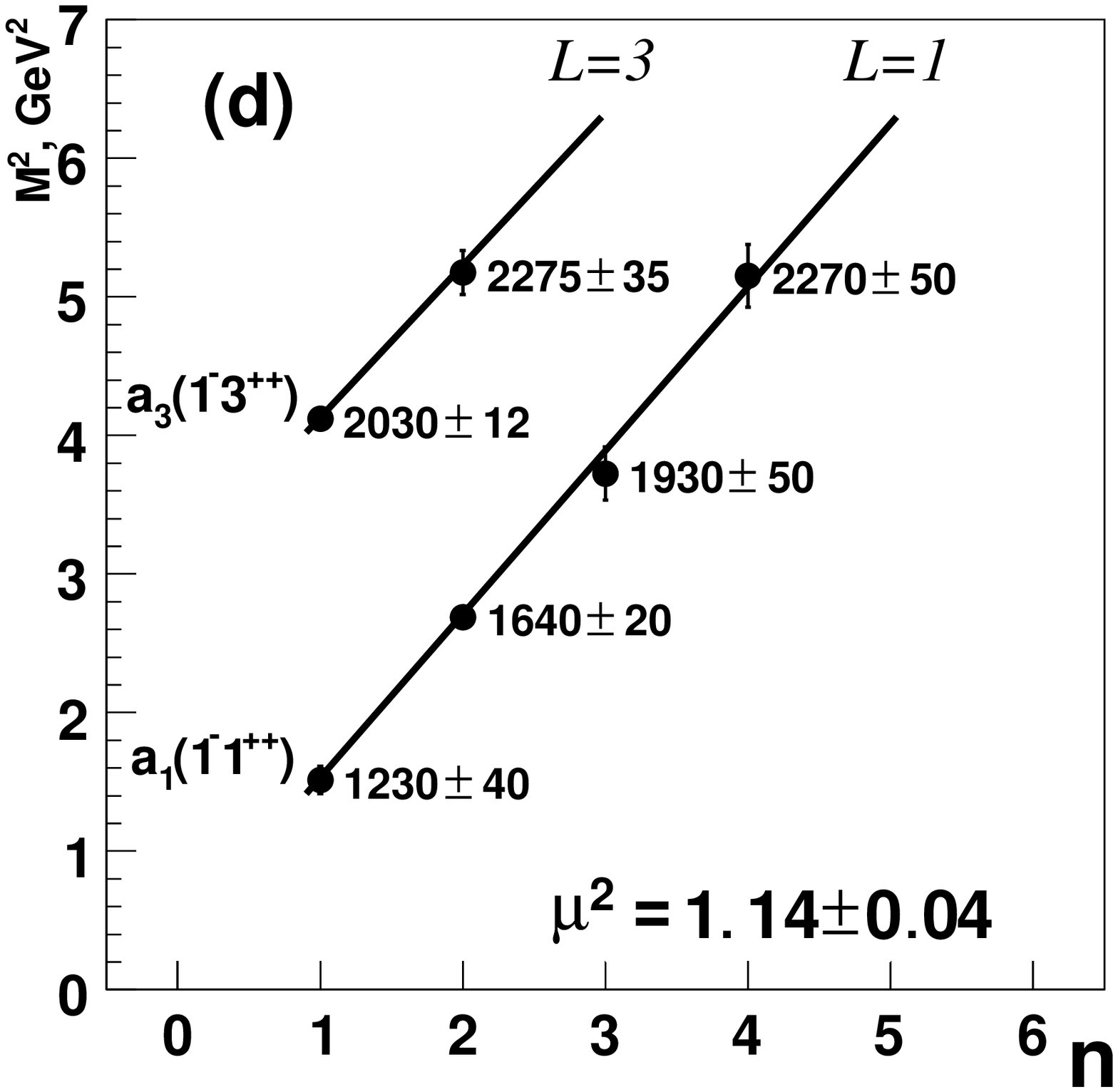,width=7.0cm}}
\vspace{-6.95mm}
\centerline{\epsfig{file=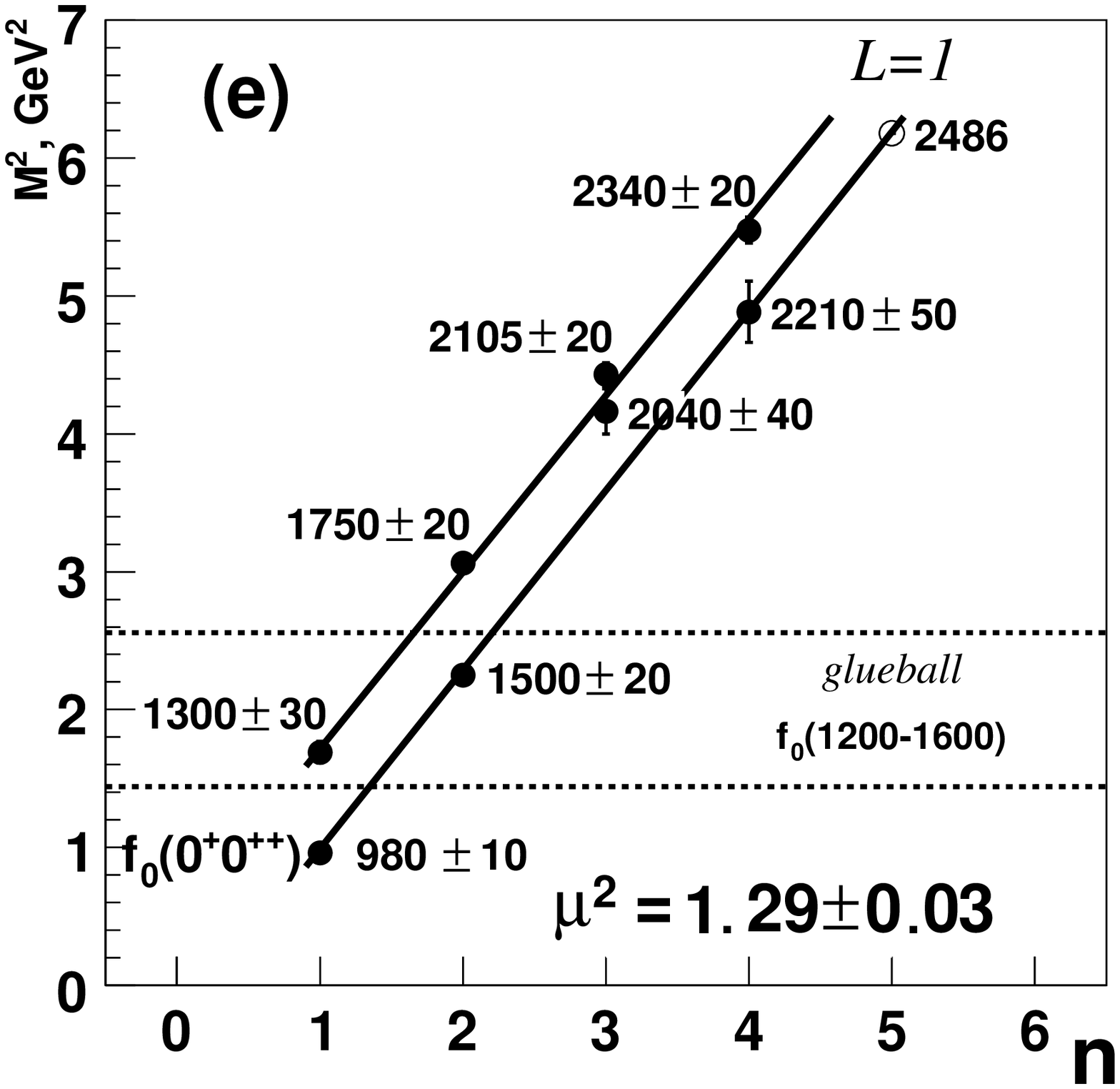,width=7.0cm}\hspace{-5mm}
            \epsfig{file=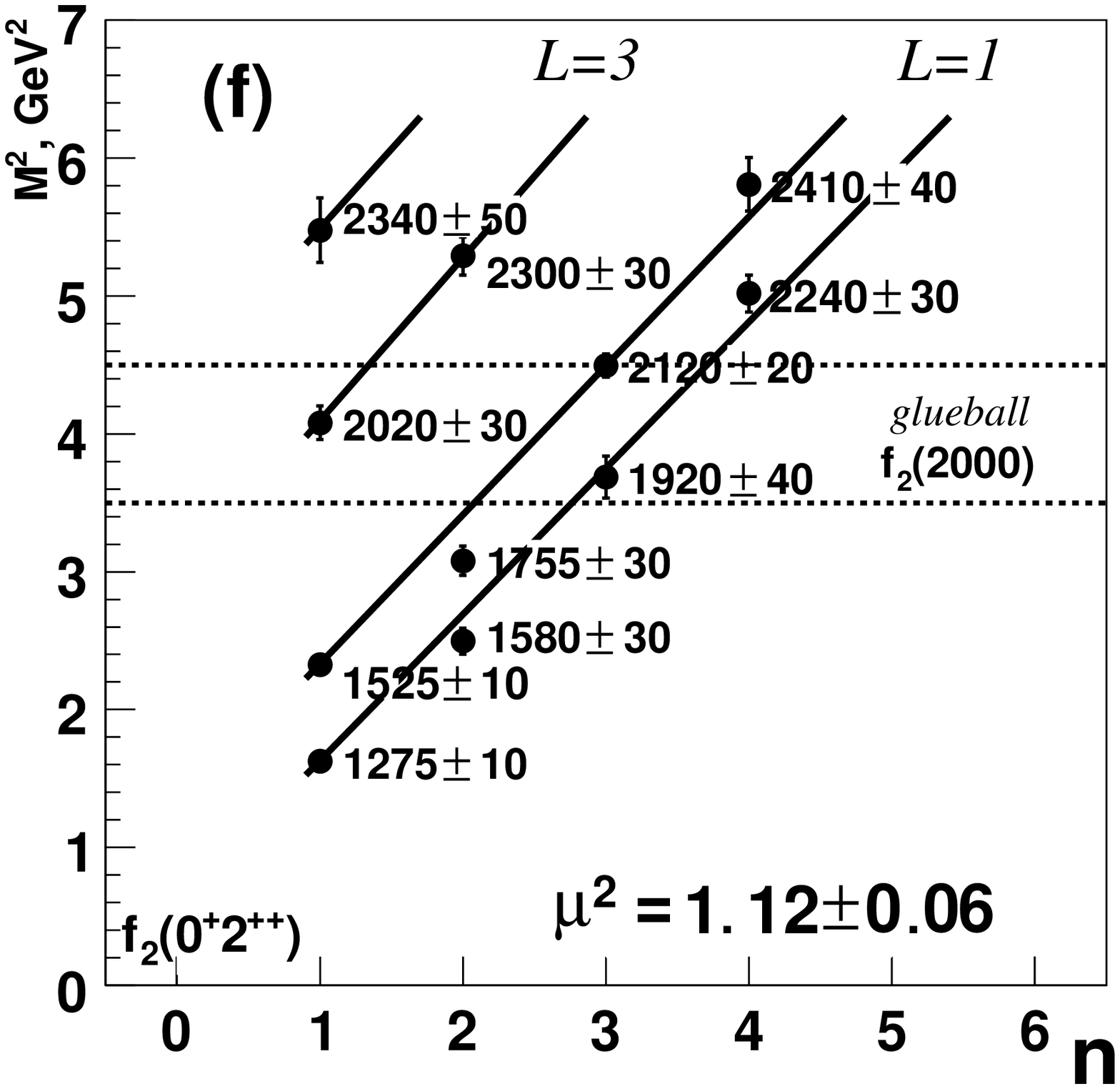,width=7.0cm}}
\vspace{-2.00mm}
\caption{Trajectories for $(C=+)$ meson states on the $(n,M^2)$ panes.
 Dotted lines
show mass regions of the scalar ($f_0(1200-1600)$) and tensor
($f_2(2000)$)  glueballs.
With the trajectories on $(n,M^2)$ plots, one can easily draw the $(J,M^2)$  ones
(Chew-Frautschi trajectories).
} \label{5}
\end{figure}

In Fig. \ref{5}, one can see states, $f_0(1200-1400)$ and $f_2(2000)$, which are extra
for the $q\bar q$ systematics. They are glueballs: scalar \cite{glueball,ufn} and tensor
\cite{gl-2} ones (hadronic decays tell us that $f_0(1200-1400)$ and $f_2(2000)$ are
nearly flavour singlet).

\section{ Spectral integral equation for $q\bar q$ mesons and confinement}
\begin{figure}
%Fig.6
\centerline{\epsfig{file=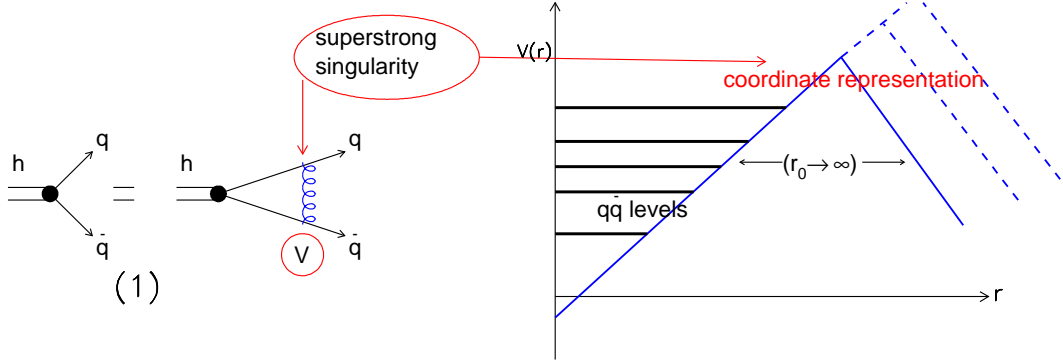,width=14cm,clip=on}}
\caption{Example which shows how in  quantum mechanics (the Schr\"odinger
equation is drawn in the left-hand side of the figure)
 one may prevent quarks to leave the confinement trap by introducing
for the $q\bar q$
system  a barrier at $r\sim r_{hadron }$ (the right-hand side of the figure).
} \label{6}
\end{figure}

In low-energy physics we
can work with constituent quarks as with standard particles but taking
into account their confinement.
 Quantum mechanics, see Fig. \ref{6}, show the way to prevent quarks to
leave the confinement trap:
for the $q\bar q$
system one should introduce a barrier at $r\sim r_{hadron}$.

The relativistic
generalization of the calculation technique was performed in terms of the
dispersion relation method \cite{sie} -- correspondingly, for $b\bar b$, $c\bar c$
and $q\bar q$ mesons.

\subsection{ Calculation results for $q \bar q$ mesons:
  masses and wave functions }

%{\epsfig{file=f3.eps,width=4cm}}  SIE

In the $q\bar q$ sector, by fitting to the interaction, there were found
wave functions and mass values of mesons lying on the following
$(n,M^2)$ trajectories:

\bea  \label{I-12}
L=0&:&\quad \pi(0^{-+}),\,
\rho(1^{--}),\, \omega(1^{--}),\, \phi(1^{--}),\\
L=1&:&\quad
a_0(0^{++}),\, a_1(1^{++}),\, a_2(2^{++}),\, b_1(1^{+-}),\,
f_2(2^{++}),\nn \\
L=2&:&\quad \pi_2(2^{-+}),
\,\rho(1^{--}),\,\rho_3(3^{--}), \,
\omega(1^{--}),\,\omega_3(3^{--}),\, \phi_3(3^{--}), \nn \\
L=3&:&\quad a_2(2^{++}),\,
a_3(3^{++}),\, a_4(4^{++}),\, b_3(3^{+-}),\, f_2(2^{++}),\,
f_4(4^{++}), \nn \\
L=4&:&\quad \rho_3(3^{--}),    \, \pi_4(4^{-+}). \nn
\eea
The linearity of trajectories on the ($n,M^2$) plane
(experimentally -- up to large $n$ values, $n\le 7$) gives us the
$t$-channel singularity:
 $V_{conf}\sim 1/q^4$ or, in the coordinate representation,
  $V_{conf}\sim  r$.
  In the coordinate representation
the confinement interaction can be written in the following
potential form:
\beq  \label{si-4}
   V_{conf}=(I\otimes I)\,b_S\,r +
(\gamma_\mu\otimes   \gamma_\mu)\,b_V\,r\ ,
\eeq
with
$b_S\simeq -b_V \simeq 0.15\,\, {\rm GeV}^{-2}$ .

The spectral integral equation for the$q\bar q$-meson wave function
was solved by introducing a cut-off into the interaction:
$r\to re^{-\mu r}$. The cut-off parameter is
small, $\mu\sim 1-10$ MeV.

\begin{figure} [h]
%Fig.7
\centerline{\epsfig{file=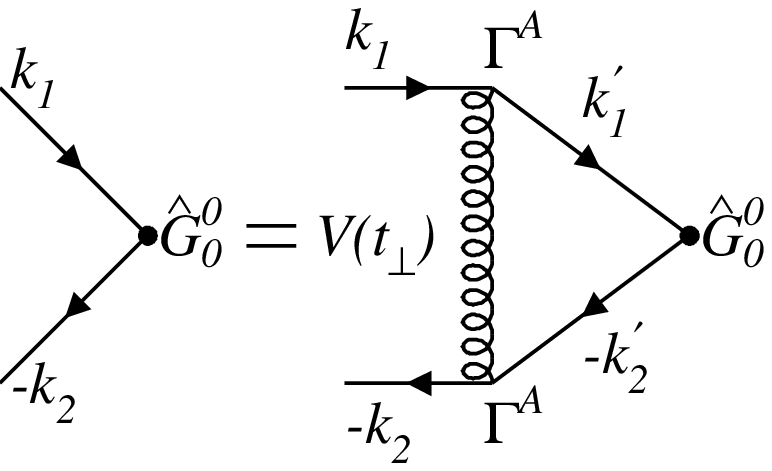,width=4cm}\hspace{18mm}
\epsfig{file=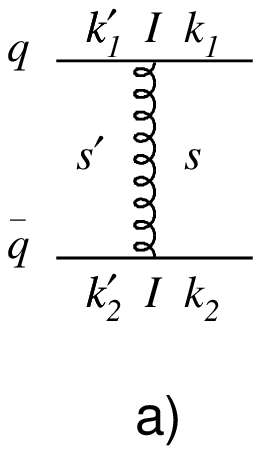,width=4cm}
}
\caption{Spectral integral equation and interaction block}\label{7}
\end{figure}

In general, keeping in mind that in the framework of spectral
integral method (as in the dispersion technique) the total energy is not
conserved, we have to write:
\beq \label{si-5}
t_{\perp}=(k^\perp_1-k^{'\perp}_1)_\mu(-k^\perp_2+k^{'\perp}_2)_\mu\ .
\eeq
 Recall that $k_1$ and $k_2$ are  momenta
of the initial quark and antiquark, while $k'_1$ and $k'_2$ are
those after the interaction. The index $\perp$ means that we use
components perpendicular to the total momentum $p=k_1+k_2$ for the
initial state and to $p'=k'_1+k'_2$ for the final state.
 \bea               \label{si-6}
&&  r^N e^{-\mu
  r}=\int\frac{d^3q}{(2\pi)^3}e^{-i\vec q\vec r} I_N(t_\perp) ,
 \quad  s=s',\; \mu \sim \frac{1}{r_0},\\
&&
  I_N(t_\perp)=\frac{4\pi(N+1)!}{(\mu^2-t_\perp)^{N+2}}
  \sum^{N+1}_{n=0}(\mu+\sqrt{t_\perp})^{N+1-n}
  (\mu-\sqrt{t_\perp})^{n}\ . \nn
\eea

\subsection{Simple example of the spectral integral equation: the pion state
($J^{PC}=0^{-+}$)}

\bea  \label{si-7}
&&\bar \psi(-k_2)i\gamma_5 g^{(0,0,0)}(s)\psi(k_1)=  \\
&&=\int\frac{d^3k'}{(2\pi)^3k_0'}\bar \psi(-k_2)V(t_\perp)
\Gamma^A(-\hat{k}_2'+m)
\frac{i\gamma_5g^{(0,0,0)}(s')}{s'-M^2}(\hat{k}_1'+m)\Gamma_A ]
\psi(k_1)
\, . \nn
 \eea
\bea
&&\hat G^{(0,0,0)}(k_\perp)=i\gamma_5,\nn \\
&&
\hat \Psi^{(0,0,0)}_{n} (s)=
\hat G^{(0,0,0)}(k_\perp)
\frac{g^{(0,0,0)}_n(s) }{s-M^2_n}=
\hat G^{(0,0,0)}(k_\perp)
\psi^{(0,0,0)}_n(s).
\eea

\section{ Two-photon decay of mesons}

Considering radiative decays of hadrons
in the spectral integral technique,
we should take into account two
subprocesses
for the two-photon decay of a $q\bar q$ state:  the
emission of a photon in the intermediate state by a quark
 (or an antiquark) and subsequant annihilation
$q\bar q\to \gamma$.

\begin{figure}
%Fig.8
\centerline{\epsfig{file=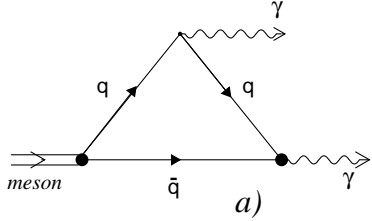,width=5.4cm}}
\caption{Two-photon decay of a meson} \label{8}
\end{figure}

\subsection{Photon wave function}

The photon wave function has two components: the soft and hard
ones. The hard component is related to the point-like vertex
$\gamma\to q\bar q$, it is responsible for the production of a
quark--antiquark pair at high photon virtuality.  The soft component is
responsible for the production of low-energy quark--antiquark vector
states such as $\rho^0$, $\omega$, $\phi(1020)$ and their excitations.

The photon wave function for
the $q\bar q$ system reads:
%\vspace{-0.2cm}
\beq
\label{SeeI2}
 \psi_{\gamma^*(Q^2)\to q\bar q}(s)=
\frac{G_{\gamma  \to q\bar q}(s)}{s+Q^2}\ .
\eeq
The photon wave function has been found
 under the assumption
 of the vertex universality
for $u$ and $d$ quarks,
$G_{\gamma\to u\bar u}(s)=G_{\gamma\to d\bar d}(s)\equiv G_{\gamma}(s)$.
 It looks rather trustworthy because of the
degeneracy trajectories  of $\rho$ and $\omega$ states.

\subsection{ Transition form factors  $\pi^0 \to \gamma
^*(Q_1^2 ) \gamma ^*(Q_2^2 )$}

In Fig. 9, one can see the results of the calculation of  transition form
factors $\pi^0,\eta,\eta'\to \gamma(Q^2)\gamma $, which appeared to be
 in a good agreement with experimental data.

Using the same technique as for the $meson \to \gamma^*(Q^2 ) V$
amplitude, we can write the formulae for the transition form factors
of the pseudoscalar mesons
$\pi^0, \eta,\eta' \to \gamma^*(Q_1^2 ) \gamma ^*(Q_2^2 )$.

The general structure of the amplitude for these processes is as
follows:
\beq
\label{Spi31}
A^{(P\to\gamma^*\gamma^*)}_{\mu\nu}(Q_1^2,Q_2^2)=
e^2\epsilon_{\mu\nu\alpha\beta}q_\alpha
p_\beta F_{(\pi,\eta,\eta')\to\gamma^*\gamma^*}(Q_1^2,Q_2^2)\ ,
\eeq
here $q=(q_1-q_2)/2$, $p=q_1+q_2$, $q_i^2=-Q_i^2$.

In terms of the light-cone  variables $(x,\vec p)$
the amplitude for the form
factor $\pi^0\to \gamma ^*(Q^2_1)\gamma ^*(Q^2_2) $  reads:
\bea
&&F_{\pi\to \gamma^* \gamma^*
}(Q_1^2,Q_2^2)= \zeta_{\pi\to \gamma \gamma }
\frac{\sqrt{N_c}}{16\pi^3} \int\limits_0^1 \frac{dx}{x(1-x)^2}\int
d^2k_\perp \Psi_{\pi}(s)
\nn \\
&&\times \left(S_{\pi\to \gamma^* \gamma^* }(s,s'_1,Q^2_1)
\frac{G_{\gamma^*}(s'_1)}{s'_1+Q^2_2}+
S_{\pi\to \gamma^* \gamma^* }
(s,s'_2,Q^2_2)\frac{G_{\gamma^*} (s'_2)}{s'_2+Q^2_1}
\right ) , \label{Spi32}
\eea
where
$s=(m^2+k_\perp^2)/[x(1-x)], $
$s'_i=[m^2+({\bf k}_\perp-x{\bf Q}_i)^2]/[x(1-x)],\, (i=1,2)$.
and
$S_{\pi\to \gamma^* \gamma^* }(s,s'_i,Q^2_i)
=4m, \qquad
\zeta_{\pi\to \gamma \gamma }=(e^2_u-e^2_d)/\sqrt 2= 1/(3{\sqrt 2})$.

 The factor $\sqrt{N_c}$  appears
owing to the definition of the colour wave function for the photon
which differs from that for the pion: in the pion wave function
there is a factor $1/\sqrt{N_c}$ while in the photon wave function
this factor is absent.

In terms of the spectral
integrals over the $(s,s')$ variables, the amplitude
for $\pi^0\to \gamma^*(Q_1^2) \gamma^*(Q_2^2)$ reads:
\bea
&& F_{\pi\to \gamma^* \gamma^* } (Q_1^2,Q_2^2)= \zeta_{\pi\to \gamma \gamma
}\frac{\sqrt{N_c}}{16}\int \limits_{4m^2}^\infty
\frac{ds}{\pi}\frac{ds'}{\pi} \Psi_\pi(s)\times
\nn \\
&&\times\left
[\frac{\Theta(s'sQ_1^2-m^2\lambda(s,s',-Q_1^2))}{\sqrt{\lambda
(s,s',-Q_1^2)}}S_{\pi\to\gamma^*\gamma^*}(s,s',Q^2_1)
\frac{G_{\gamma^*} (s')}{s'+Q^2_2}
\right.
\nn \\
&&\left.+
\frac{\Theta(s'sQ_2^2-m^2\lambda(s,s',-Q_2^2))}{\sqrt{\lambda
(s,s',-Q_2^2)}}S_{\pi\to\gamma^*\gamma^*}(s,s',Q^2_2)
\frac{G_{\gamma^*} (s')}{s'+Q^2_1}
\right ] ,
\label{Spi35}
\eea
where $\lambda(s,s',-Q_i^2)=(s-s')^2+2Q_i^2(s+s')+Q_i^4$,
$\quad\Theta(X>0)=1$ and $\Theta(X<0)=0$.

\begin{figure}
%Fig. 9
\centerline
{\epsfig{file=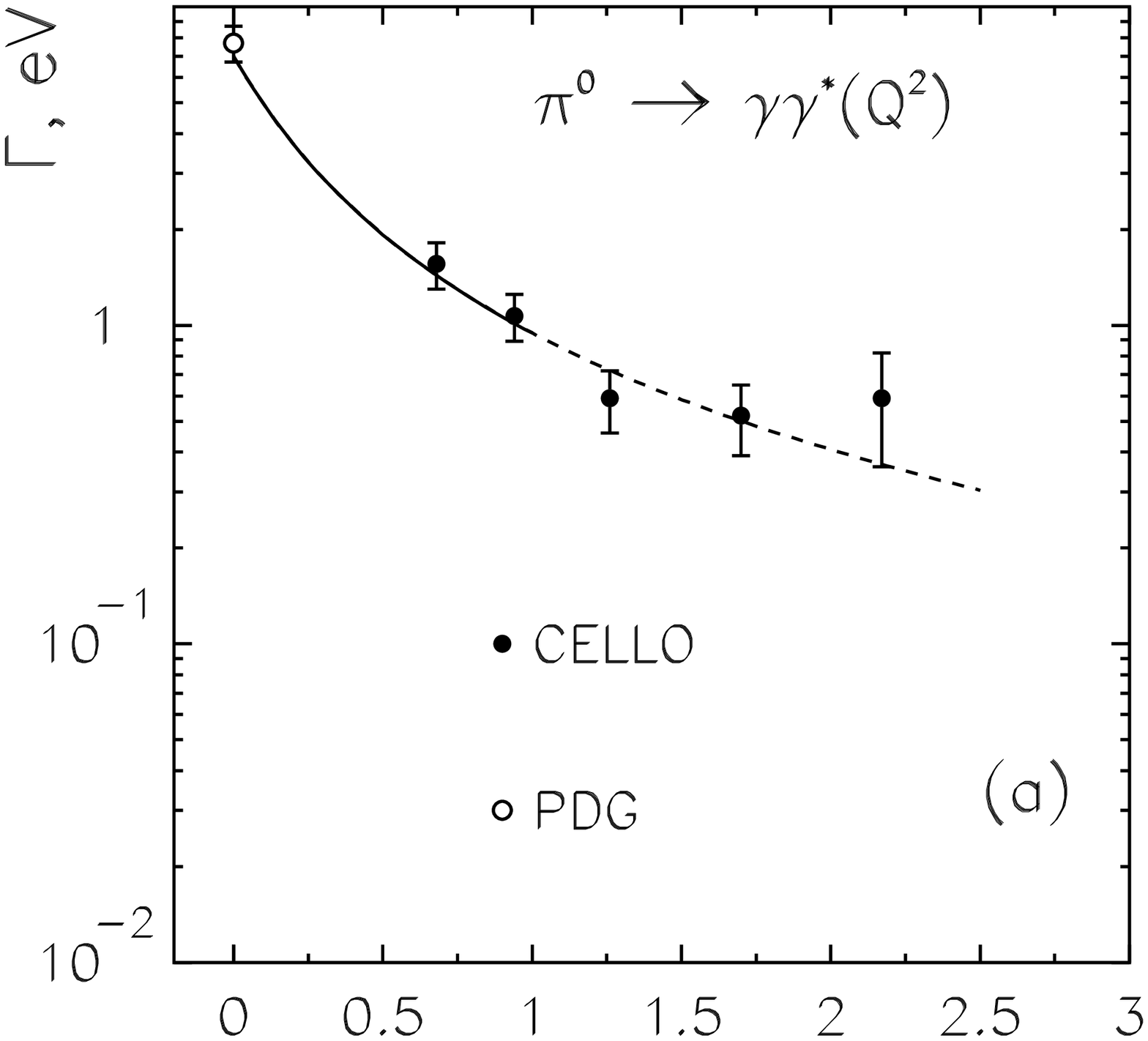,width=0.3\textwidth}
            \epsfig{file=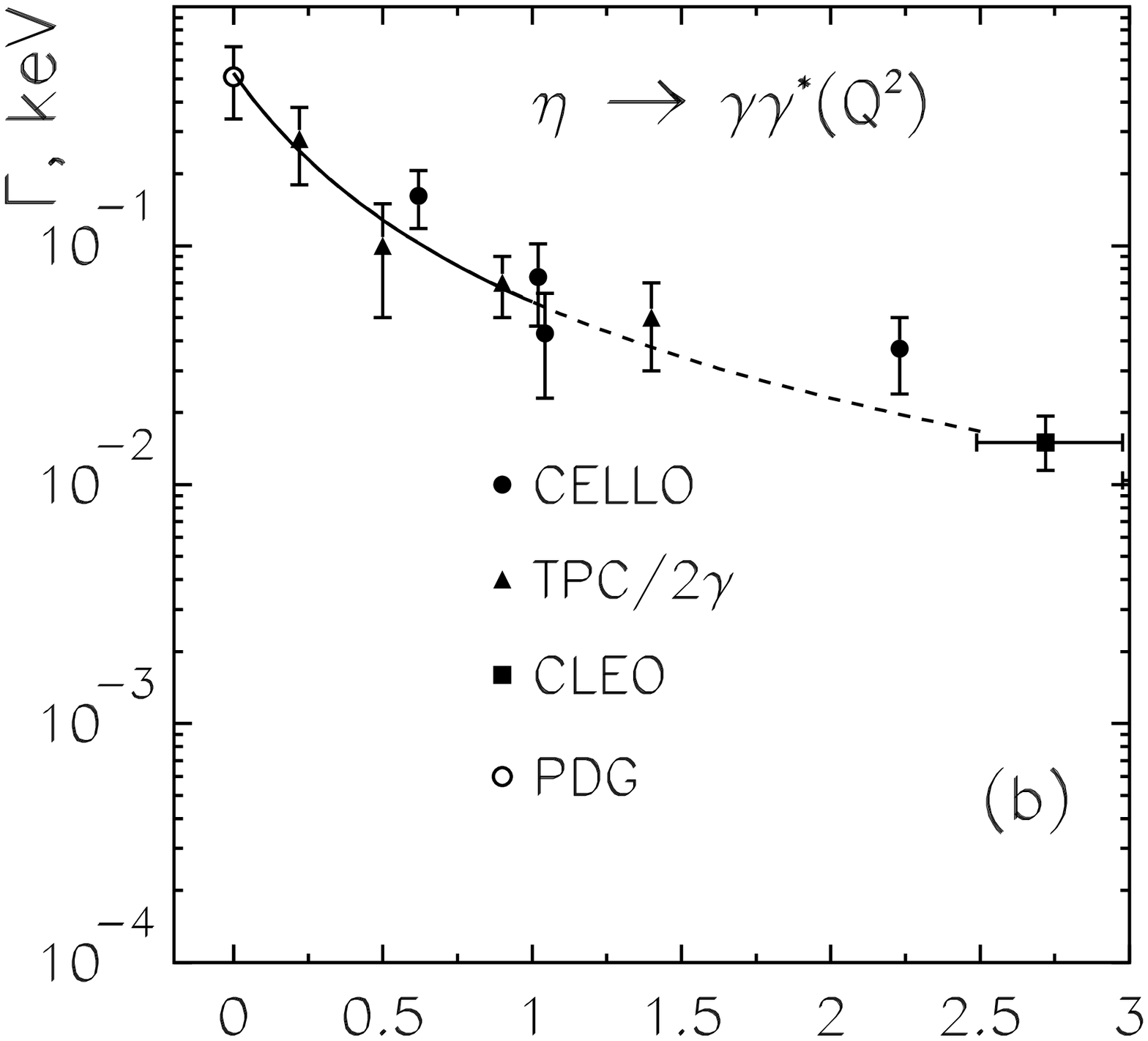,width=0.3\textwidth}
\epsfig{file=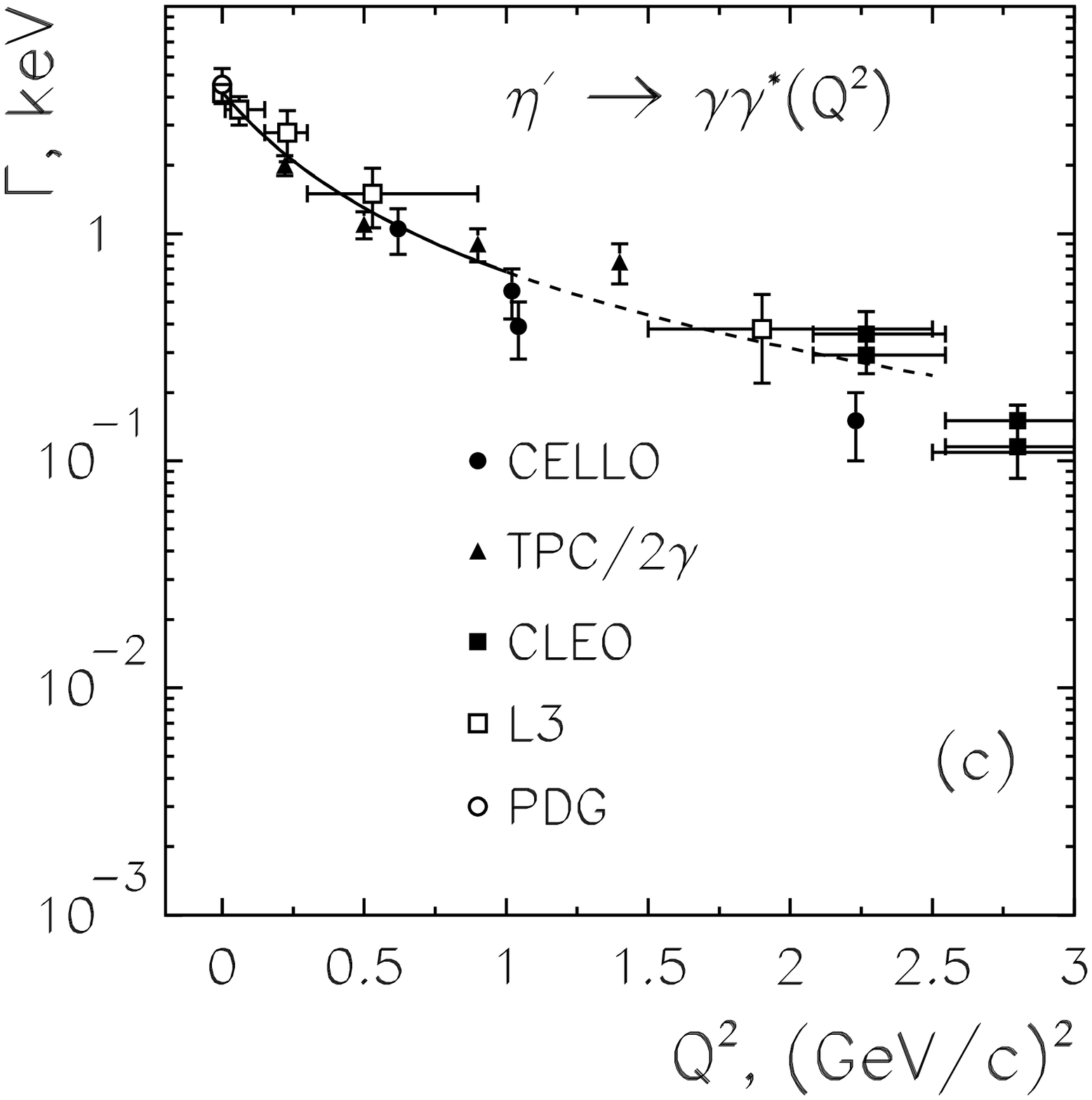,width=0.3\textwidth}}
\caption{ Data for
$\pi^0\to\gamma\gamma^*$,
$ \eta\to\gamma\gamma^*$,
$\eta'\to\gamma\gamma^*$
{\it vs}
 calculated curves.}\label{9}
\end{figure}

In the radiative decays $\rho,\omega\to\gamma\pi$ we face two
mechanisms: a bremsstrahlung emission of a photon $q\to \gamma+q$, with
a subsequent transition $q\bar q\to \pi$, and
a bremsstrahlung-type emission of a pion $q\to \pi+q$, with
a subsequent annihilation $q\bar q\to \gamma$.

\begin{figure}
%Fig. 10
\centerline{\epsfig{file=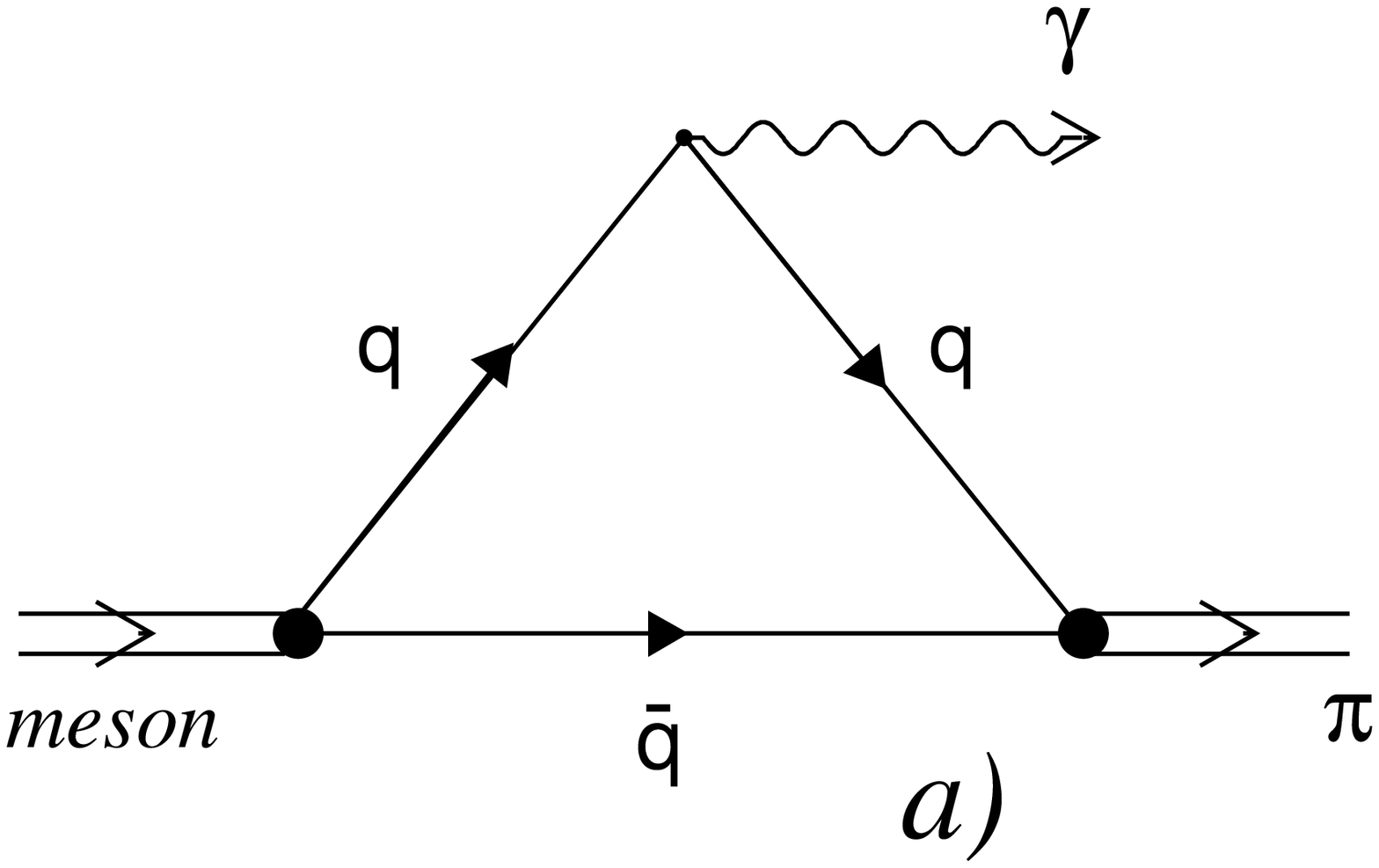,width=4.0cm}
\epsfig{file=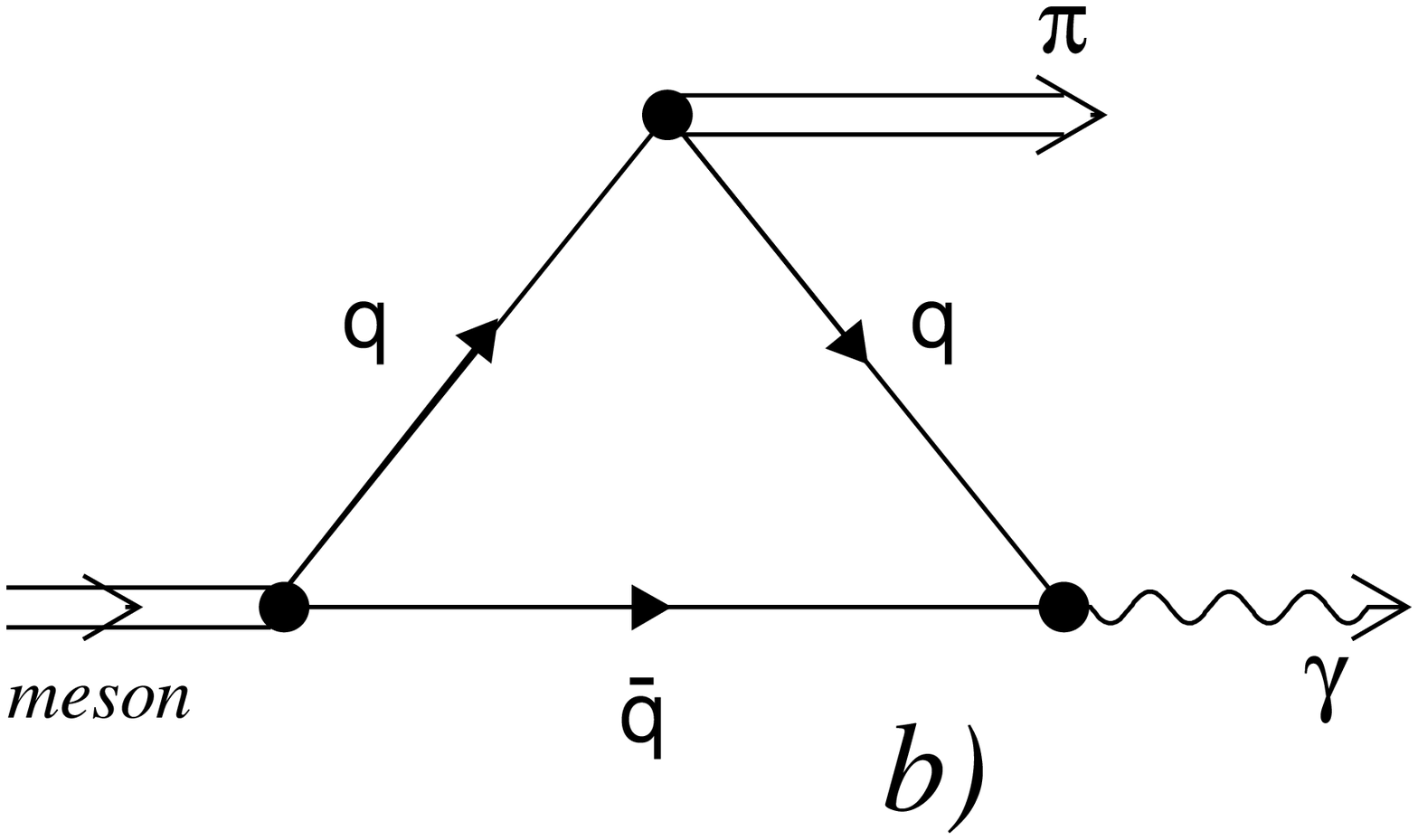,width=4.0cm}}
\caption{The $\rho,\omega\to\gamma\pi$ decay}\label{10}
\end{figure}

The key point in
the calculation of the $\rho,\omega\to \gamma\pi$ decays is to know the
$q\bar q$ wave functions of the pion and vector mesons, $\rho$ and
$\omega$, as well as the wave function of the  photon
$\gamma\to q\bar q$.
The pion bremsstrahlung constant for the process $q\to \pi q$ is
determined by the pion-nucleon coupling constant
 $g_{\pi\pi N}^2/(4\pi)\simeq 14$.
We have calculated
partial widths
-- they coincide with the observed ones:
$\Gamma^{(exp)}_{\rho^\pm\to \gamma\pi^\pm}= 68\pm 30$ keV,
$\Gamma^{(exp)}_{\rho^0\to\gamma\pi^0}= 77\pm 28$ keV,
$\Gamma^{(exp)}_{\omega\to\gamma\pi^0}= 776\pm 45$ keV.

\section{ Decay $\rho\to \pi\pi$}

\begin{figure}
%Fig. 11
\centerline
{\epsfig{file=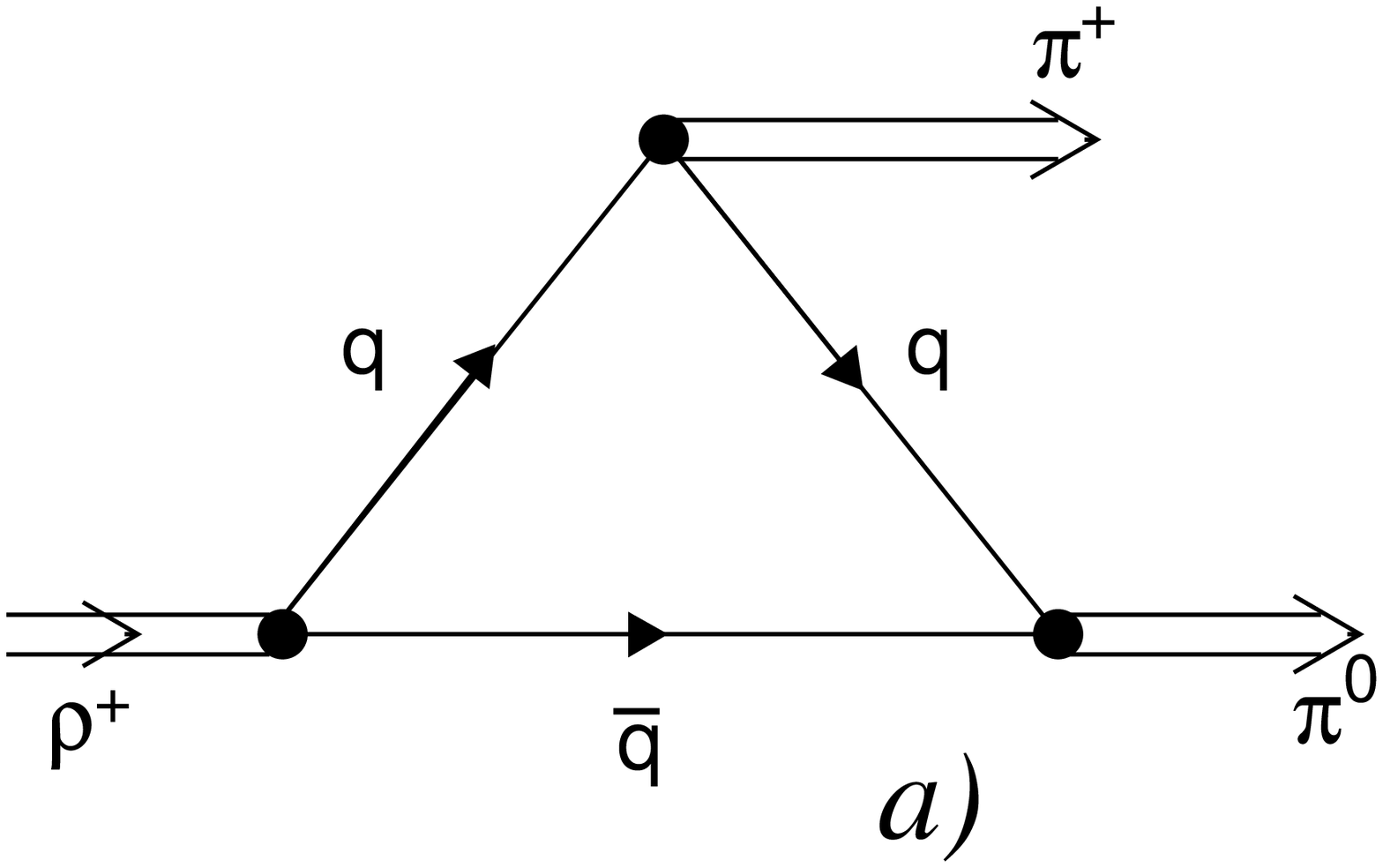,width=4.0cm}
\epsfig{file=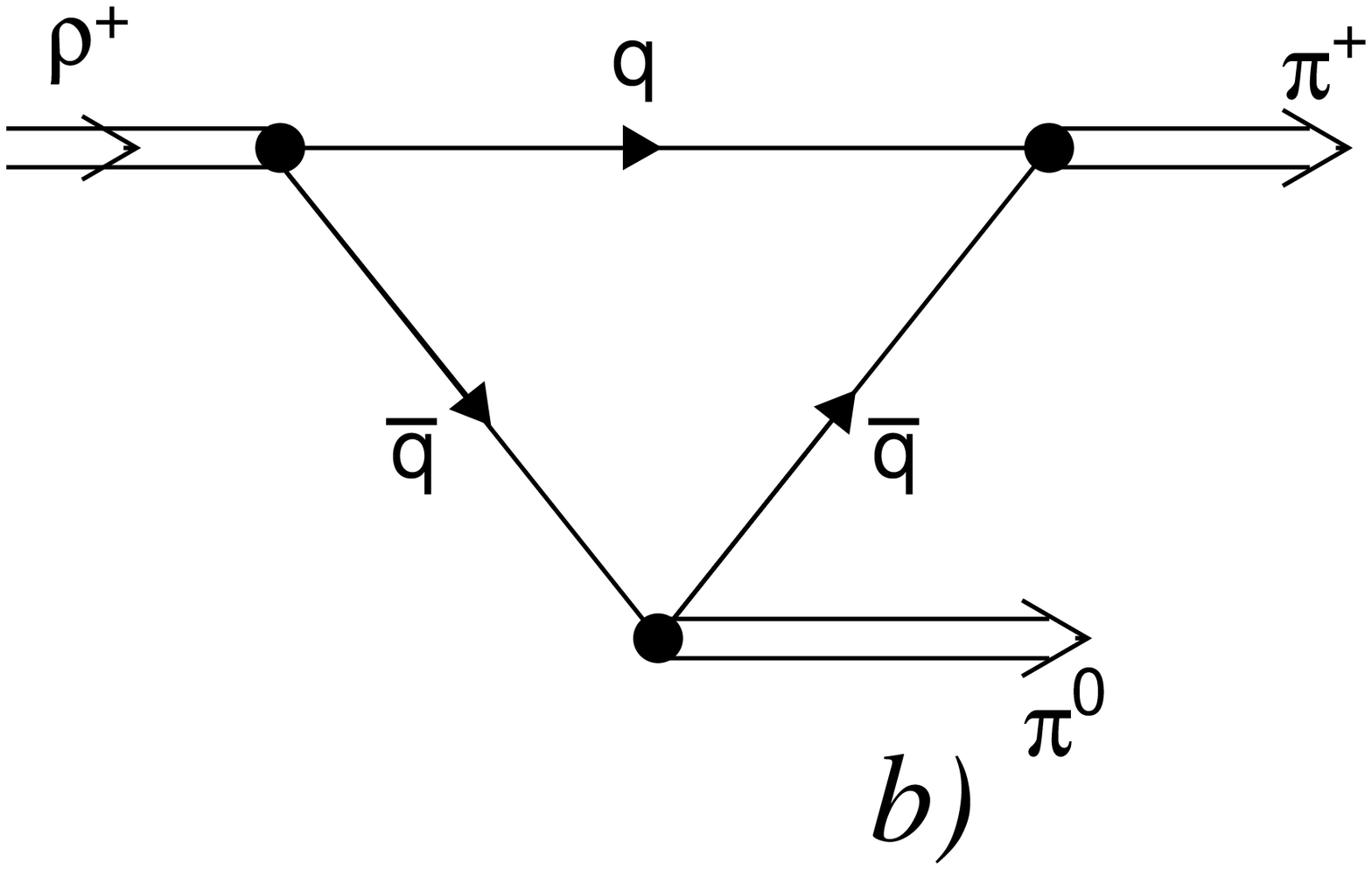,width=4.0cm}}
\caption{ Decay $\rho\to\pi\pi$.
These two diagrams, being taken into account only, result in
$\Gamma_{\rho(775)\to\pi\pi} \simeq 2000$ MeV}\label{11}
\end{figure}

The study of $\rho\to \pi\pi$
 demonstrates us that in this decay, apart from the pion
bremsstrahlung processes, we should include
into consideration the Gribov  quark exchange \cite{Gribov}.

 \begin{figure}
%Fig.12
\centerline{\epsfig{file=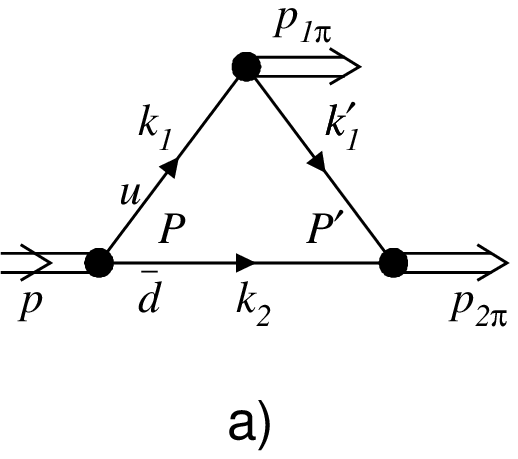,width=3.7cm}
               \epsfig{file=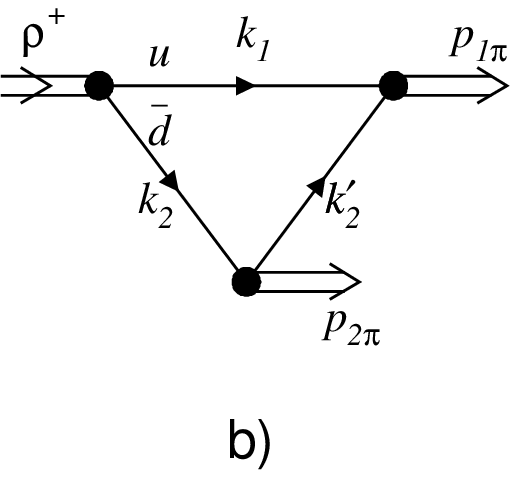,width=3.7cm}
                 \epsfig{file=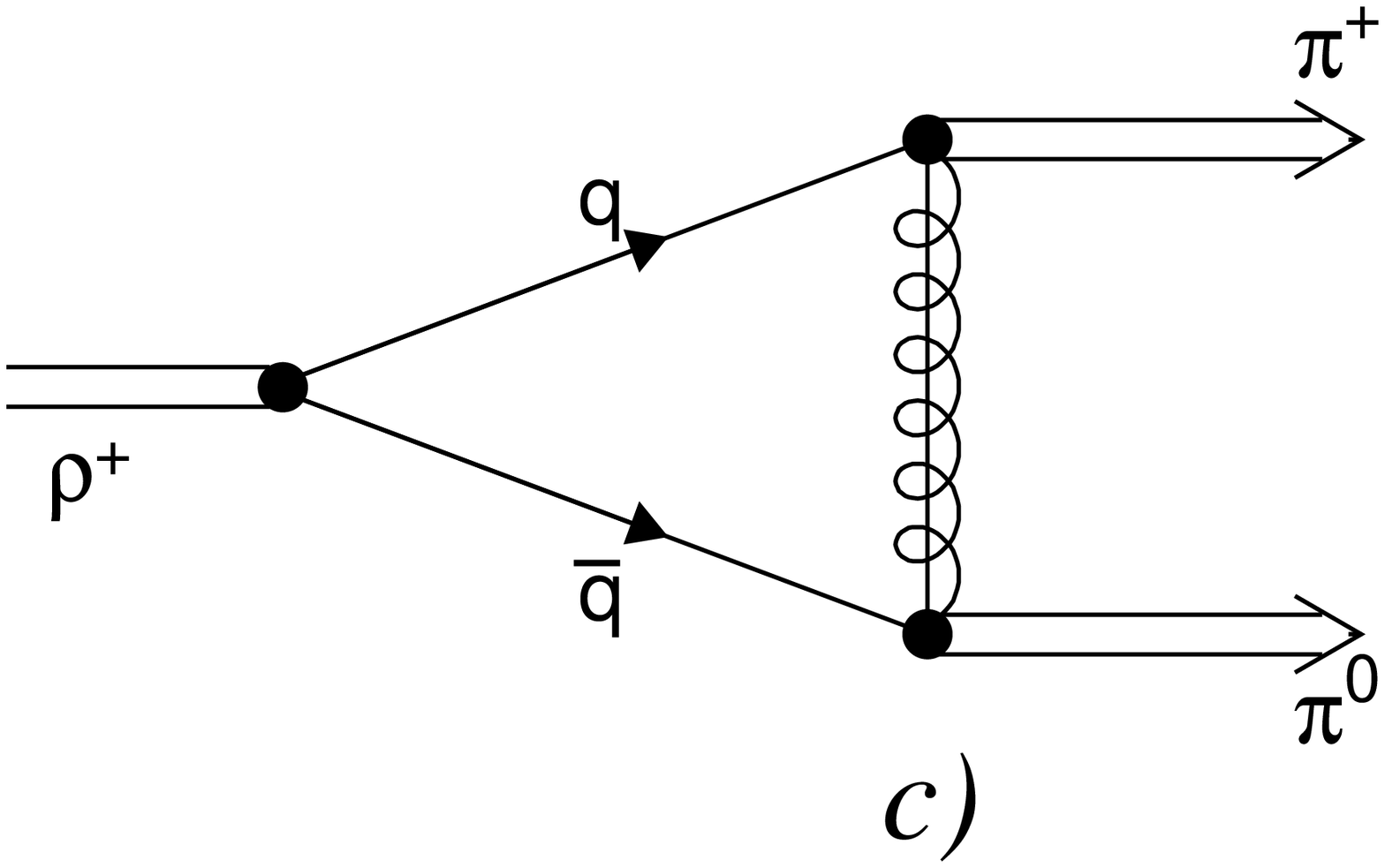,width=4.2cm}}
\caption{
 The $\rho^+\to\pi^+\pi^0$ process: a,b) with a
  bremsstrahlung-type pion emission and c) with the $q\bar q\to\pi\pi$
  transition, realized owing to the $t$-channel confinement
  singularity with fermion quantum numbers.
These three diagrams give $\Gamma_{\rho(775)\to\pi\pi} \simeq 150$
MeV}\label{12}
\end{figure}

The Gribov's transition  $q\bar q\to \pi\pi $, see Fig. \ref{12}c,  is necessary for
the numerical description of the $\rho(775)\to\pi\pi$ decay width.
However,
the structure of the $q\bar q\to \pi\pi$ amplitude is not
determined unambiguously -- there remains a freedom in choosing the
singularity.

The estimate of the amplitude $q\bar q\to \pi\pi$ for
different interaction versions has been done in \cite{conf}.

\begin{figure}
%Fig 13
\centerline
{\epsfig{file=f7a.eps,width=4cm,clip=on}
    \epsfig{file=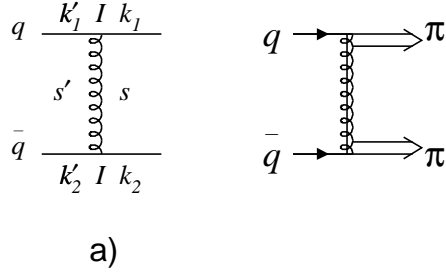,height=4cm}}
\caption{Confinement interactions: singular locking block and Gribov's quark exchange.}
\label{13} \end{figure}

The calculation of amplitudes demonstrates that for  the
quark confinement  principally decisive is the   singular
interaction: $1/t^2_\perp$ -- just because of this
interaction the quark singularities are absent in the decay amplitudes.
Correspondingly,
the amplitudes are  real in the physycal region.

The Gribov's singularity
plays another and very important role. The matter
is that the bremsstrahlung-type radiation of pions ({\it i.e.}  the
radiation coming from the region of the confinement trap) is rather
large. Were it the only possible process, it would lead to a
broad decay width of the $\rho$ meson, $\sim 2000$ MeV. But the process
with  Gribov's singularity  prevents such a ``smearing'', -- it
looks like the only way for keeping the $\rho$ meson as a comparatively
narrow state.

\subsection{ Self-energy part  $\rho\to\pi\pi\to\rho$}

\begin{figure}
%Fig. 14
\centerline{\epsfig{file=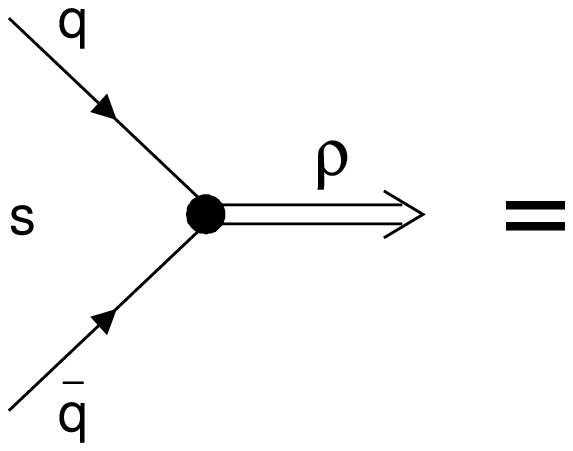,height=3cm}
            \epsfig{file=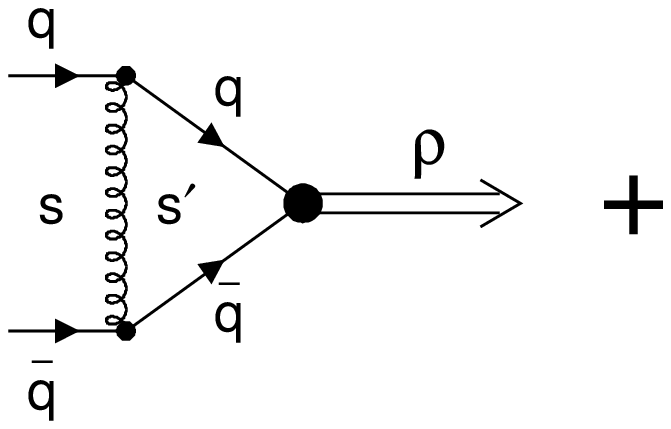,height=3cm}
            \epsfig{file=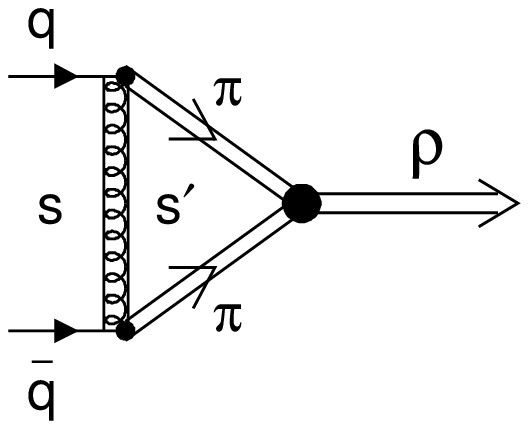,height=3cm}}
\vspace{-5mm}
\centerline{\epsfig{file=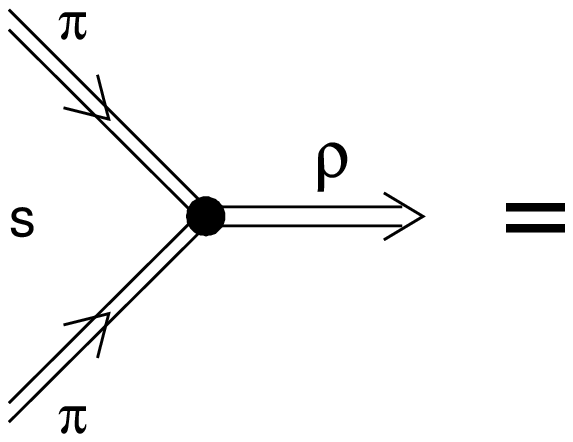,height=3cm}
            \epsfig{file=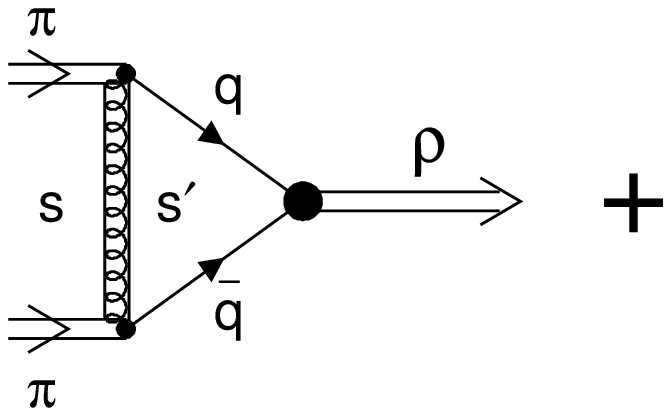,height=3cm}
            \epsfig{file=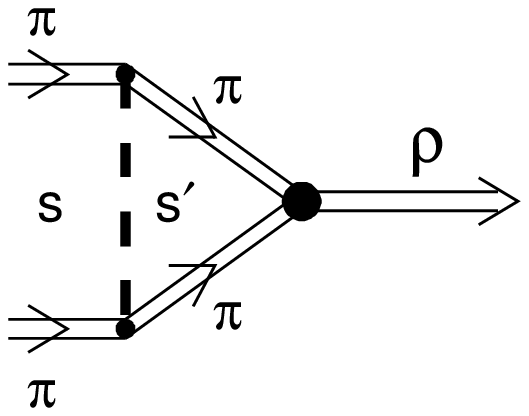,height=3cm}}
\vspace{-5mm}
\centerline{\epsfig{file=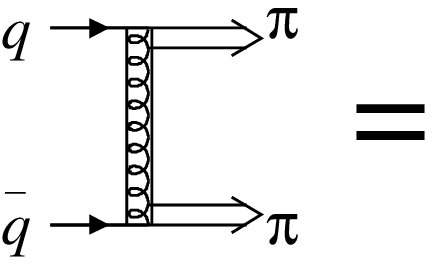,height=3cm}
            \epsfig{file=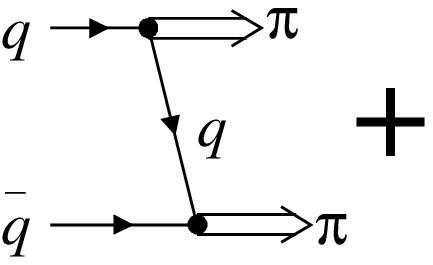,height=3cm}
            \epsfig{file=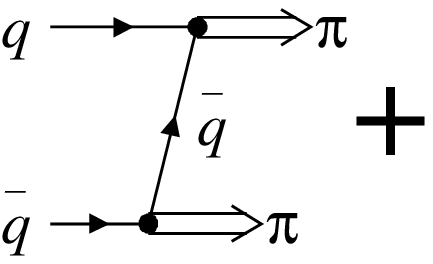,height=3cm}
            \epsfig{file=f11d.eps,height=3cm}}
\vspace{-5mm}
\caption{ Two-channel
($q\bar q$, $\pi\pi$) equation:
 The following simplifying steps can be done:
(i) The  second term in the right-hand side of the
($q\bar q\to \rho$)-equation may be considered as a perturbative
 correction because the $\rho$-meson width is not large.
(ii) The last term in the equation $\pi\pi \to \rho$ can be neglected
due to the smallness
of the $\pi\pi$ interaction in the $\rho$-meson region.}
\label{14}
\end{figure}

Up to now
we neglect the reverse influence of the
$\pi\pi$ channel upon the characteristics of the $\rho$ meson. To
take it into account one needs to consider the two-channel
($q\bar q$, $\pi\pi$) equation, see Fig. \ref{14}.

\begin{figure}
%Fig 15
\centerline{\epsfig{file=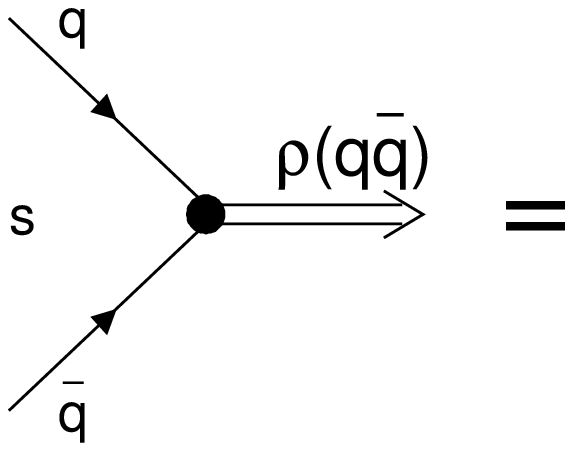,height=3.4cm}
            \epsfig{file=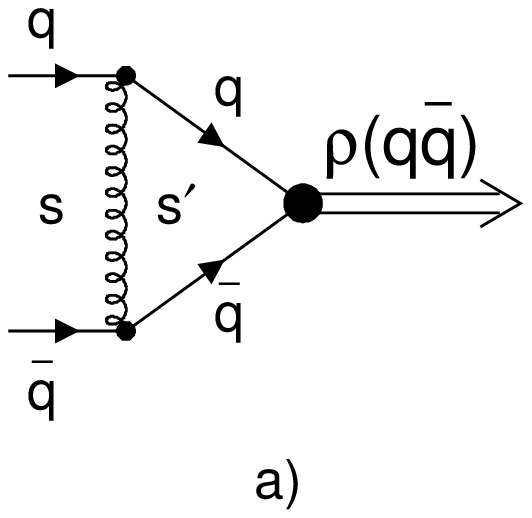,height=3.4cm}}
\vspace{-5mm}
\centerline{\epsfig{file=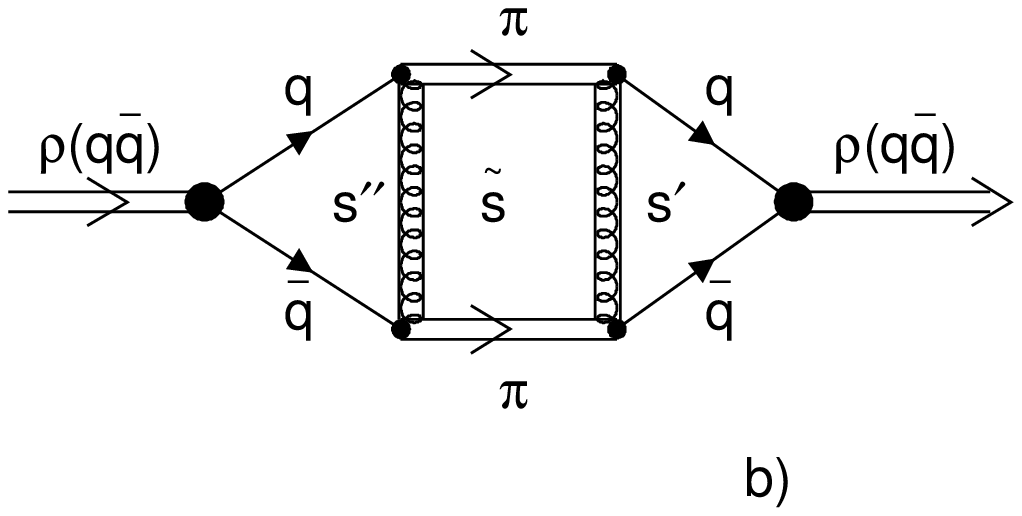,height=3.4cm}}
\caption{Standard
$q\bar q$ equation and the $\rho$-meson
self-energy part
 $\rho_{(q\bar q)}\to \pi\pi\to \rho_{(q\bar q)}$.
}
\label{15}
\end{figure}

A small value of the ($\rho_{(q\bar q)}\to \pi\pi$) width allows us to simplify
the equation of Fig. \ref{14}  -- see Fig. \ref{15}. Then, we have a standard one-channel
equation for $\rho_{(q\bar q)}$ where $\rho_{(q\bar q)}$ is a pure $q\bar q$ state. The
$\pi\pi$ channel reveals itself in the self-energy part
 $\rho_{(q\bar q)}\to \pi\pi\to \rho_{(q\bar q)}$.

 One can see that the self-energy part $B(s,M^2_{\rho(q\bar q)})$ determines the
 admixture of
the $\pi\pi$ component in the $\rho$ meson, and the amplitude has hadron singularities
only; quark cuttings give zero contributions.

Indeed, with this self-energy part,
$B(s,M^2_{\rho(q\bar q)})$, one transforms
 the propagator of a pure $q\bar q$
state:
\bea \label{G4}
&&\frac{\sum\limits_a \epsilon^{(a)}_\nu
\epsilon^{(a)+}_{\nu'}}{M^2_{\rho(q\bar q)}-s}
\quad \Rightarrow\quad
\frac{\sum\limits_a \epsilon^{(a)}_\nu
\epsilon^{(a)+}_{\nu'}}{M^2_{\rho(q\bar q)}-s-B(s,M^2_{\rho(q\bar q)})}=\\
&&= \frac{\sum\limits_a \epsilon^{(a)}_\nu \epsilon^{(a)+}_{\nu'}}
{M^2_{\rho(q\bar q)}-{\rm Re}B(s,M^2_{\rho(q\bar q)})
-s-i\,{\rm Im}B(s,M^2_{\rho(q\bar q)})}
\simeq \frac{\sum\limits_a \epsilon^{(a)}_\nu \epsilon^{(a)+}_{\nu'}}
{M^2_{\rho}-s-i\,M_{\rho}\Gamma_{\rho}}\ ,  \nn
\eea
where
\bea \label{G5}
 &&
M^2_{\rho}= M^2_{\rho(q\bar q)}-{\rm Re}\,B(M^2_{\rho},M^2_{\rho(q\bar q)})
= (0.770)^2{\rm GeV}^2 ,\\
&&
 M_{\rho}\Gamma_{\rho}={\rm Im}\,B(M^2_{\rho},M^2_{\rho(q\bar q)}).
\nn
\eea
The fit  tells us that
$M^2_{\rho}\simeq M^2_{\rho(q\bar q)}$, thus providing us
with a rather small value of the mass shift
in the region of  $s\sim M^2_{\rho}$:
\beq \label{G6}
{\rm Re}\,B(M^2_{\rho},M^2_{\rho(q\bar q)})\simeq 0.
\eeq

\subsection{ Self-energy part  $B(s,M^2_{\rho(q\bar q)})$ and
confinement}

The self-energy part of a pure $q\bar q$ state reads:
% {\epsfig{file=f12d.eps,height=2.4cm}}

\bea
 &&B(s,M^2_{\rho(q\bar q)})=\int\limits_{4M^2_\pi}^\infty \frac{d
s_{\pi\pi}}{\pi}\frac{{\rm Im}B(s_{\pi\pi},M^2_{\rho(q\bar q)})}{
s_{\pi\pi}-s-i0}, \\
&&
{\rm Im}B( s_{\pi\pi},M^2_{\rho(q\bar
q)})
=\frac{1}{48\pi} \sqrt{\frac{( s_{\pi\pi}-4M^2_\pi)^3}{ s_{\pi\pi}}}
[ 2A
(\bigtriangleup^{\pi^+}_{\pi^0}; s_{\pi\pi},M^2_\rho)+
A(\triangleleft^{\pi^+}_{\pi^0}; s_{\pi\pi},M^2_\rho)
]^2 .\nn
\eea
We should emphasize that the self-energy amplitude,
despite the
presence of the quark--antiquark state in intermediate states,
 have no  corresponding imaginary part.
This means the quark confinement.
The only particles
flying away
are pions: the threshold singularity in the amplitude
 manifests  just
this fact.

\section{  Conclusion}

The equations for $q\bar q$ mesons, which are constructed in terms
of dispersion relation technique (or spectral integral  technique),
are in certain respect similar to
Schr\"odinger equations with infinite potential barrier.  Spectral
integral eqations, taking into account the
relativism of  constituents, make it
possible to describe the confined $q\bar q$ systems and
to calculate  meson masses and $q\bar q$ wave functions.
For the quark confinement,  principally crucial are   singular
interactions \cite{book3}.

Singular
interactions, imposing the confinement trap constraints, allow us to
calculate the radiative decay amplitudes and widths:
$(q\bar q)_{in}\to\gamma^*\to e^+ e^-$,
$(q\bar q)_{in}\to\gamma\gamma$, $(q\bar q)_{in}\to\gamma (q\bar q)_{out}$.

A special treatment is needed for reactions with the production of
$\gamma\pi$ and two-pion decay. Here we face the
bremsstrahlung-type emission of pions ({\it i.e.},  the
emission coming from the region of the confinement trap).
Besides, in the two-pion production, $\rho\to\pi\pi$, a special type of
singular interaction  must  exist, namely,
Gribov's singularity -- it
prevents a ``smearing'' of the $\rho$ meson.

So, the calculations tell us that in terms of the dispersion
relation technique we need three types of diagrams for the description of the transition
$q\bar q\to \pi\pi$:
with the emision
of a pion by quark,  antiquark, and a
  simultaneous production of two pions that
 plays a role of a subtraction term in the interaction block.

The calculation of the self-energy part $\rho \to \pi\pi \to \rho$
demonstrates that the model of constituent quarks, in its standard
version presented here, does not leave a room
for the inclusion of the
pion cloud into consideration. Indeed, along with
pion cloud component, it is necessary to decrease the size and mass of
the ''$q\bar q$ bag''
($R^2_{q\bar q-{\rm bag}}<R^2_{\rho -{\rm meson}}$),
{\it i.e.}, it is necessary to change the logic of
spectral integral equations for mesons.

 I thank L.G. Dakhno and M.A. Matveev for useful comments.
%\clearpage

\end{document}